\documentclass{aa}
\usepackage{amsmath,graphics,amssymb}

\usepackage{natbib}
\bibpunct{(}{)}{,}{a}{}{,}

\newcommand{\Teff}{\ensuremath{T_\mathrm{eff}}}
\newcommand{\zav}[1]{\left(#1\right)}
\newcommand{\ms}{\ensuremath{\text{M}_{\odot}}}

\newcommand{\msr}{\ensuremath{\ms\,\text{yr}^{-1}}} 
\newcommand{\vel}{{v}}
\newcommand{\vinfty}{\ensuremath{\vel_\infty}}
\newcommand{\hzav}[1]{\left[#1\right]}

\newcommand\de{\text{d}}

\allowdisplaybreaks

\begin{document}

\title{CMF models of hot star winds}
\subtitle{II. Reduction of O star wind mass-loss rates in global models}

\author{J.  Krti\v{c}ka\inst{1} \and J. Kub\'at\inst{2}}
\authorrunning{J. Krti\v{c}ka and J. Kub\'at}

\institute{\'Ustav teoretick\'e fyziky a astrofyziky P\v{r}F MU,
            CZ-611 37 Brno, Czech Republic, \email{krticka@physics.muni.cz}
           \and
           Astronomick\'y \'ustav, Akademie v\v{e}d \v{C}esk\'e
           republiky, CZ-251 65 Ond\v{r}ejov, Czech Republic}

\date{}

\abstract{We calculate global (unified) wind models of main-sequence, giant, and
supergiant O stars from our Galaxy. The models are calculated by solving
hydrodynamic, kinetic equilibrium (also known as NLTE) and comoving-frame (CMF)
radiative transfer equations from the (nearly) hydrostatic photosphere to the
supersonic wind. For given stellar parameters, our models predict the
photosphere and wind structure and in particular the wind mass-loss rates
without any free parameters. Our predicted mass-loss rates are by a factor of
2--5 lower than the commonly used predictions. A possible cause of the
difference is abandoning of the Sobolev approximation for the calculation of the
radiative force, because our models agree with predictions of CMF NLTE radiative
transfer codes. Our predicted mass-loss rates agree nicely with the mass-loss
rates derived from observed near-infrared and X-ray line profiles and are
slightly lower than mass-loss rates derived from combined UV and H$\alpha$
diagnostics. The empirical mass-loss rate estimates corrected for clumping may
therefore be reconciled with theoretical predictions in such a way that the
average ratio between individual mass-loss rate estimates is not higher than
about $ 1.6 $. On the other hand, our predictions are by factor of $ 4.7 $ lower
than pure H$\alpha$ mass-loss rate estimates and can be reconciled with these
values only assuming a microclumping factor of at least eight.}


\maketitle

\section{Introduction}

Hot massive stars lose a significant amount of their mass via line-driven winds
\citep{lusol,cak}. The mass loss influences stellar evolution
\citep{china,dedela,kostel} and contributes to the mass and momentum input into
the interstellar medium. One of the key parameters that determine the influence
of the stellar wind on stellar evolution and on the circumstellar medium is the
amount of mass lost per unit of time, that is, the mass-loss rate.

There are several observational indicators that are used to derive the wind
mass-loss rates. The shape of X-ray emission line profiles is affected by the
continuum absorption in cool wind \citep{macawl,tlustocerv,lidarikala} providing
the the possibility to determine the mass-loss rates. The strength of the
ultraviolet P~Cygni lines also scales with the mass-loss rate, but the
sensitivity of P~Cygni lines to the ionization balance complicates their usage
for the mass-loss rate determination \citep{lamoc,full}. H$\alpha$ emission line
is another suitable mass-loss rate tracer \citep{pulchuch} that can be most
easily reached. Near-infrared emission lines \citep{najaro} and the excess at
long wavelengths \citep{scupamoc,dalijaty} are also used to probe the wind
mass-loss rate.

Hot star wind mass-loss rates can be also predicted using dedicated wind models.
The models typically use kinetic equilibrium equations (statistical equilibrium
equations) to derive the ionization and excitation equilibrium, however the
models differ in the treatment of the hydrodynamics and radiative transfer.
While \citet{vikolamet} apply the Monte Carlo method of the solution of the
radiative transfer equation and prespecified velocity law, \citet{graham,
grahamz} and \citet{cmf1} use comoving frame (CMF) radiative transfer equation
and solve hydrodynamics self-consistently.

Ideally, all observational tracers should give the same mass-loss rate for the
same star, which should also agree with theoretical predictions. However, this
is not the case in most cases, and the differences may amount to one order of
magnitude \citep[e.g.,][]{pulamko}. This is a serious problem for stellar
evolutionary models and for the determination of the massive star feedback,
for example, the influence of massive star on the interstellar medium. The discordant
mass-loss rate determinations may be possibly reconciled by taking into account
the effect of small scale wind inhomogeneities (clumping) on the diagnostics and
on the predictions. These inhomogeneities are most likely connected with
intrinsic instability of the line driving \citep{ocr,felpulpal,opsim,runow}. The
instabilities may be possibly triggered by photospheric turbulence
\citep{cant,jian}.

The small scale wind inhomogeneities (discussed in the proceedings
\citealt{potclump}) are typically divided into optically thin ones
(microclumping) and optically thick ones (macroclumping or porosity). The
optically thin inhomogeneities affect the wind properties that are proportional
to the square of the density. Consequently, they affect the H$\alpha$ emission
line, which originates due the the recombination, and free-free emission
resulting from the collisions of protons with free electrons
\citep[e.g.,][]{pulchuch}. Microclumping affects also the wind ionization
equilibrium and therefore the strength of ultraviolet wind line profiles
\citep{prvnifosfor,bezchuch}. The inhomogeneities may be optically thick in
continuum, possibly affecting the shape of X-ray line profiles
\citep{lidarikala,oskirec}, and optically thick in lines, affecting the
ultraviolet wind line profiles \citep{osporcar}. To proceed toward more
realistic mass-loss rates on the observational side, it is necessary to
understand the influence of optically thin and optically thick inhomogeneities
on the observational tracers \citep{chuchcar,sund,clres1,clres2,shendelori}.

Because the small scale wind inhomogeneities affect the ionization balance and
the radiative transfer, they may also influence the theoretical wind mass-loss
rate predictions. The inhomogeneities influence only the terminal wind velocity
if they start at large distances from the star \citep{sanya,supujo}, whereas the
inhomogeneities that start very close to the photosphere affect also the wind
mass-loss rate \citep{sanya,muij}. Moreover, to understand the reasons for a
possible discordance between theoretical and observational mass-loss rate
estimates, all approximations in the theoretical models have to be carefully
inspected.

One of the most important simplifications that appeared in our previous models
\citep{cmf1} was the so-called core-halo approximation. This approach separates
treatment of the atmosphere (core) and the wind (halo). Within this
approximation, we accounted for the influence of the stellar radiation on the
radiative force and wind ionization equilibrium. However, the influence of wind
on atmosphere was neglected. Models that provide more realistic connection of
the wind and the photosphere abandoning the core-halo approximation (unified or
global models) have been calculated for O stars
\citep{acko,gableri,hilmi,pahole,puluni,ssh}. However, only the models of
\citet{grahamz} and \citet{powrdyn} allow also for selfconsistent treatment of
the radiative force from the (nearly) hydrostatic photosphere to supersonic wind

Here we study the effect of abandoning the core-halo approximation on predicted
mass-loss rates. We have included the photosphere into our models and use the same
hydrodynamic and radiative transfer equations both in the photosphere and in the
wind. We also improved the kinetic equilibrium equations. The models newly
account for the influence of line transitions on the radiation field, however
they still rely on the Sobolev approximation during the calculation of
bound-bound radiative rates.

\section{Global wind models}

The wind models were calculated using the METUJE code which solves equations
describing stellar photosphere and wind together. The code solves the radiative
transfer equation, the kinetic equilibrium equations, and hydrodynamic equations
in photosphere and wind, and enables us to properly describe the interaction of
photospheres and winds. The code solves stationary (time-independent) equations
assuming spherical symmetry.

The model calculation consists of two loops, the inner loop and the outer loop
(c.f., \citealt{nltei}, Fig.~1). The inner loop (Sect.~\ref{nlte}) solves the
NLTE problem, that is, the kinetic equilibrium equations (with Sobolev
bound-bound term) and continuum radiative transfer equation (with corrections
for lines) to obtain consistent values of level populations and radiation field.
The inner loop is iterated until the convergence is achieved. The outer loop
solves the comoving frame (CMF) radiative transfer equation (Sect.~\ref{cmf})
and performs iterations of hydrodynamic equations (Sects.~\ref{conmom} and
\ref{tep}) with level populations taken from the inner loop. The full inner loop
is performed for each iteration of the outer loop. The outer loop is iterated
until the hydrodynamic structure converges. The model
is
calculated for
different lower boundary velocities until the code finds the boundary velocity
which allows the solution to smoothly pass through the critical
point\footnote{The critical point in our calculations 
corresponds
to
radiative-acoustic waves \citep{abb,thofe}. It 
appears
for supersonic velocities
larger than the classical speed of sound.}. Moreover, at the beginning of the
calculation, the model 
is
calculated only for subcritical velocities. In
subsequent steps, when the lower boundary velocity is known with sufficient
precision (roughly about 50\%), the solution 
is
extended to the whole wind and
comprises also supercritical velocities.

The initial estimate of the structure of global models in the subsonic part (for
velocities lower than the classical speed of sound) was derived from the TLUSTY
atmosphere models \citep{ostar2003} calculated for the same stellar parameters
(stellar effective temperature \Teff, surface gravitational acceleration $g$,
and chemical composition) as the global model. The method for obtaining the initial
estimate of the structure of global models in the supersonic part is described
by \citet{nltei}. Both parts of the initial estimate are continuously connected
to enable smooth run of iterations.

\subsection{Radiative transfer equation}
\label{cmf}

The radiative transfer equation 
is
solved numerically as a second order
differential equation in the comoving frame (CMF) using the method of
\citet{mikuh}. We 
use
a modified version of the original method. As in our
previous paper \citep{cmf1}, our method 
specifies
the flux-like variable
$v_\nu(\tau_\nu,\mu)=(I_\nu(\tau_\nu,\mu)- I_\nu(\tau_\nu,-\mu))/2$ at the grid
points \citep{cmf1}. The frequency grid \citep[see][for details]{cmf1} is
constructed with spacing $\Delta\nu_{\text{D},l}$ proportional to the thermal
speed of a fictitious atom with a mass of $m_\text{C}=60\,m_\text{H}$
($m_\text{H}$ is a mass of the hydrogen atom) at the temperature
$T_\text{C}=20\,$kK,
$\Delta\nu_{\text{D},l}=\nu_l\sqrt{\zav{2kT_\text{C}/m_\text{C}}}/
\zav{cf_\text{D}}$, where $l$ is the grid index, and $f_\text{D}=2$. Typically
we use about $10^6$ frequencies to solve the CMF radiative transfer equation.
For wind velocities smaller than the thermal speed the CMF radiative transfer
equation approaches the static radiative transfer equation, consequently the
same CMF equation can be used to model both photosphere and wind.

We 
account
for all relevant processes that influence the radiative transfer in
hot star winds, that is, the bound-bound and bound-free processes of considered
ions \citep [for their list see][] {nlteiii}, free-free processes of \ion{H}{i},
\ion{He}{i}, and \ion{He}{ii}, and scattering on free electrons. All these
processes are included in the absorption and emission coefficients. The emission
coefficient describing the scattering on free electrons 
is
calculated using
continuum mean intensity corrected for lines, see Eq.~\eqref{uziteJ} below. For
our calculations we 
assume
Gaussian line profiles and thermal broadening only.
The line (bound-bound) data used for the calculation of opacity and emissivity
were extracted from the VALD database (Piskunov et al. \citeyear{vald1}, Kupka
et al. \citeyear{vald2}). Moreover, we tested an extended line list including 41
million lines of \ion{Fe}{iii}\,--\,\ion{Fe}{vi} taken from the Kurucz
website\footnote{\url{http://kurucz.harvard.edu}}. The tests showed that the
resulting line force differs by just few percent from that calculated using
shorter line list. Consequently, we adopted the original shorter line list, for
which the calculations are faster.

The inner boundary condition for the radiative transfer equation was derived from
the diffusion approximation \citep[see][Chapter~11.9]{hubenymihalas}. We write
the diffusion approximation in terms of the source function
\citep[see][]{auerhermite}, which provides smoother models than the Planck
function. The first three terms of the Taylor expansion of the source function
are
\begin{equation}
\label{staylor}
S_\nu(t_\nu)=S_\nu(\tau_\nu)+\frac{\de S_\nu}{\de\tau_\nu}(t_\nu-\tau_\nu)
+\frac{1}{2}\frac{\de^2 S_\nu}{\de\tau_\nu^2}(t_\nu-\tau_\nu)^2,
\end{equation}
where $\tau_\nu$ is the radial optical depth.
The substitution to the formal solution of the static radiative transfer
equation gives \citep[see][Eq.~11.165]{hubenymihalas}
\begin{equation}
I_\nu(\tau_\nu,\mu)=S_\nu(\tau_\nu)+\mu\frac{\de S_\nu}{\de\tau_\nu}+
\mu^2\frac{\de^2 S_\nu}{\de\tau_\nu^2}.
\end{equation}
This yields the inner boundary condition
\begin{equation}
v_\nu(\tau_\nu,\mu)=\mu\frac{\de S_\nu}{\de\tau_\nu},
\end{equation}
or, in the finite difference approximation, as implemented in the code,
\begin{multline}
v_1(\nu)=\frac {\mu}
{\chi_1(\nu)}\left[\frac{S_1(\nu)-S_3(\nu)}{r_3-r_1}-\right.\\
\left.-2\zav{\frac{S_2(\nu)-S_3(\nu)}{r_3-r_2}-\frac{S_1(\nu)-S_2(\nu)}{r_2-r_1}
}\frac{r_2-r_1}{r_3-r_1}\right].
\end{multline}
To obtain this equation, we extrapolated the finite differences to estimate the
difference at the boundary. The subscripts number the grid points. Numerical
tests have shown that inclusion of higher-order terms in Eq.~\eqref{staylor}
does not improve the precision of the solution. The outer boundary condition
\citep{cmf1} assumes no infalling radiation.

\subsection{Kinetic equilibrium equations}
\label{nlte}

The level populations are derived from the kinetic equilibrium equations
\citep[see] [Chapter~9] {hubenymihalas}. We take into account all relevant
atomic processes important in hot star winds, namely the radiative and
collisional excitation, deexcitation, ionization, recombination, and Auger
processes. The radiative field for the calculation of radiative rates
corresponding to continuum transitions is taken from the solution of the CMF
radiative transfer equation. To speed up the NLTE iterations, the bound-bound
terms are derived using the Sobolev method \citep{rybashumrem} with incident
intensity ($I_\text{c}$ in Eq.~(33) of \citealt{rybashumrem}) derived from the
CMF radiative field. As a good approximation we took the radiation at the outer
model boundary. The tests we carried out using different incident intensities (corresponding to
points closer to the photosphere) showed that this approximation does not
significantly affect the resulting general model structure.

The kinetic equilibrium equations have to be solved together with the radiative
transfer equation to obtain the level populations that are consistent with
radiative field. The whole process of solution must be iterated to get the
converged solution in each step of the iteration of hydrodynamical equations.
However, the CMF solution of the radiative transfer equation is time consuming
and its execution after each kinetic equation iteration step would require
prohibitively long computation time. To avoid this, we 
solve
the CMF radiative
transfer equation only before the iteration step of hydrodynamic equations, and
use the CMF solution to correct the mean intensity derived from continuum
radiative transfer equation. The continuum radiative transfer equation
is
calculated neglecting line transitions \citep{nltei}. The solution of the
continuum equation is therefore much faster and can be calculated after each
solution of the kinetic equilibrium equations. To correct the mean intensities
$J_\nu^\text{cont}$ derived from the continuum radiative transfer equation for
the effect of line transitions, we use the correction factors \citep[blocking
factors, c.f.,][]{pahole} calculated as
\begin{equation} \label{slozityzlomek}
d_\nu^\text{CMF}=\frac{\bar{J}_\nu^\text{CMF}}{J_\nu^\text{cont}},
\quad\text{or}\quad d^\text{CMF}(\nu_i)=\frac{1}{\Delta\nu_{i}}
\frac{\int\limits_{\overline\nu_i}^{\overline\nu_{i+1}}{J}_\nu^\text{CMF}
\,\de\nu}
{J_{\nu_i}^\text{cont}} .
\end{equation}
These factors are derived using correct mean intensities from the solution of
CMF radiative transfer equation ${J}_\nu^\text{CMF}$ averaged over frequencies
corresponding to each $J_{\nu_i}^\text{cont}$. Here
$\Delta\nu_{i}=(\nu_{i+1}-\nu_{i-1})/2$, $\overline\nu_{i}=
(\nu_i+\nu_{i-1})/2$, $\nu_{i}$ are frequencies of a grid at which
$J_\nu^\text{cont}$ is evaluated, and $i$ is a frequency index. For simplicity,
we omitted the radial dependence in Eq.~\eqref{slozityzlomek}. The correction
factors $d_\nu^\text{CMF}$ are recalculated using Eq.~\eqref{slozityzlomek}
after each CMF solution. Then the mean intensity, which is used in kinetic
equilibrium equations, is calculated as
\begin{equation}\label{uziteJ}
J_\nu = d_\nu^\text{CMF} J_\nu^\text{cont}
\end{equation}
from each solution of the continuum radiative transfer equation.

Similarly as in the case of the radiative force correction factors
$c^\text{CMF}$ (see \citealt{cmf1}) the direct use of factors $d^\text{CMF}$
defined by Eq.~\eqref{slozityzlomek} causes instability in the model
convergence. To avoid this we introduced weak smoothing of $d^\text{CMF}(\nu)$
for $2 < j < \mathrm{NR}-1$ (NR is the number of radial points),
\begin{equation}
\label{vyhlad}
\overline
d^\text{CMF}_j(\nu_i)=\frac{1}{4}\hzav{2d^\text{CMF}_j(\nu_i)+
d^\text{CMF}_{j-1} (\nu_i)+d^\text{CMF}_{j+1}(\nu_i)},
\end{equation}
where $d^\text{CMF}_j (\nu_i)$ is the value of $d^\text{CMF} (\nu_i)$ at a given
grid point $j$ (as for $j-1$ and $j+1$) and we use $\overline d^\text{CMF}_j
(\nu_i)$ instead of $d^\text{CMF}_j(\nu_i)$ in the models. Our numerical tests
showed that the smoothing Eq.~\eqref{vyhlad} does not significantly affect the
global properties of the model but eliminates unphysical point to point
variations of the temperature in the atmosphere.

To accelerate the iterative solution of the radiative transfer equation together
with the kinetic equilibrium equations (the NLTE problem) we use accelerated
lambda iterations based on the method proposed by \citet{rybashumremali}. The
linearization of derived equations is based on Newton-Raphson iterations
resembling the method of \citet{hemso}. This enabled us to naturally combine the
acceleration of iteration due to continuum transitions with Newton-Raphson
iterations derived from the Sobolev method used for the line transitions
\citep{nlteiii}.

The kinetic equilibrium equations are also used to calculate the derivatives of
the occupation numbers with respect to flow variables, most notably the
temperature (see Sect.~2.4 of \citealt{nltei}). These derivatives are used
within the Newton-Raphson solution of hydrodynamic equations. Contrary to
previous models, we added an additional term that accounts for the dependence of
the mean intensity on the temperature in Eq.\,(24) of \citet{nltei}. This term
is calculated assuming that the derivative of the mean continuum intensity of
radiation $J_\nu^\text{cont}$ is the same as in LTE, that is,
\begin{equation}
\label{djdt}
\frac{\partial J_\nu^\text{cont}}
{\partial T}=\frac{1}{1-\exp\zav{-\frac{h\nu}{kT}}}
\frac{h\nu J_\nu^\text{cont}}{kT^2}.
\end{equation}
This additional term is used only deep in the atmosphere in the regions, where
the temperature is derived from the differential form of the transfer equation
(see Eq.~\eqref{integ}).

The model ions for the solution of the kinetic equilibrium equations were either
adopted from TLUSTY model atmosphere input files \citep [see][for their
description] {ostar2003,bstar2006} or prepared by us using Opacity and Iron
Project data \citep{topt,zel0} and data described by \citet{pahole}. The list of
included ions is given in \citet{nlteiii}.

\subsection{Continuity and momentum equations}
\label{conmom}

The continuity and momentum equations \citep[Eqs.~(20.1) and
(20.21)]{hubenymihalas} are the same in the photosphere and in the wind.
However, their physical interpretations differ. The momentum equation with the
radiative force term enables us to derive the radial velocity in the wind,
whereas in the limiting case of the stellar photosphere this equation approaches
the hydrostatic equation, which is used for determination of density.
Consequently, the continuity equation determines the density in the wind,
whereas it is used to derive the velocity in the quasistatic photosphere.
Despite these differing physical interpretations, the model smoothly passes from
the photosphere in hydrostatic equilibrium to the supersonic wind, because we
used the same equations throughout the model.

The radiative force in the momentum equation is derived from the solution of CMF
radiative transfer equation. We included both the line radiative force and the
radiative force due to continuum transitions (most notably the light scattering
by free electrons). Our experience shows that the continuum radiative force can
significantly affect the density distribution of subsonic part of the model.
This can spoil the convergence of the model, consequently we limit the maximum
change of the continuum radiative force between subsequent iterations to a
fraction (typically about $0.1$) of the radiative force due to the light
scattering on free electrons. Moreover, we limited the maximum continuum radiative
force to twice the radiative force due to the light scattering on free
electrons.

\subsection{Equation of temperature}
\label{tep}

The equation for the temperature comprises the most difficult part of the
unification of the wind and photosphere models. This is connected with the fact
that the physical mechanisms that determine temperature in the photosphere and
in the wind are very different. Consequently, the equation used to derive the
temperature is the only equation that significantly differs in the photosphere
and in the wind.

\newcommand\ts{\times}
\begin{table*}
\caption{Adopted parameters of the model grid (stellar effective temperature
$\Teff$, radius $R_{*}$, mass $M$, and luminosity $L$), mass-loss rates $\dot
M_\text{Vink}$ derived using \citet{vikolamet} formula for \citet{asp09} mass
fraction of heavier elements, and our predicted terminal velocity $v_\infty$ and
mass-loss rate $\dot M$}
\centering
\label{ohvezpar}
\begin{tabular}{rccrccccc}
\hline
\hline
& Model &$\Teff$ & $R_{*}$ & $M$ & $\log(L/L_\odot)$ & $\dot M_\text{Vink}$ &
$v_\infty$ & $\dot M$ \\
& & $[\text{K}]$ & $[\text{R}_{\odot}]$ &
$[\text{M}_{\odot}]$ & & $[\text{M}_{\odot}\,\text{yr}^{-1}$] &
[$\text{km}\,\text{s}^{-1}$] & $[\text{M}_{\odot}\,\text{yr}^{-1}$]\\
\hline Main
&300-5& 30000 &  6.6 & 12.9 & 4.50 & $1.3\ts10^{-8}$ & 2440 & $1.0\ts10^{-8}$ \\ sequence
&325-5& 32500 &  7.4 & 16.4 & 4.74 & $4.2\ts10^{-8}$ & 1900 & $1.4\ts10^{-8}$ \\
&350-5& 35000 &  8.3 & 20.9 & 4.97 & $1.2\ts10^{-7}$ & 1220 & $3.8\ts10^{-8}$ \\
&375-5& 37500 &  9.4 & 26.8 & 5.19 & $3.0\ts10^{-7}$ & 1770 & $1.1\ts10^{-7}$ \\
&400-5& 40000 & 10.7 & 34.6 & 5.42 & $7.3\ts10^{-7}$ & 1960 & $2.5\ts10^{-7}$ \\
&425-5& 42500 & 12.2 & 45.0 & 5.64 & $1.6\ts10^{-6}$ & 1910 & $4.5\ts10^{-7}$ \\
\hline Giants
&300-3& 30000 & 13.1 & 19.3 & 5.10 & $1.6\ts10^{-7}$ & 2070 & $6.2\ts10^{-8}$ \\
&325-3& 32500 & 13.4 & 22.8 & 5.25 & $3.7\ts10^{-7}$ & 1470 & $1.2\ts10^{-7}$ \\
&350-3& 35000 & 13.9 & 27.2 & 5.41 & $8.0\ts10^{-7}$ & 1200 & $2.4\ts10^{-7}$ \\
&375-3& 37500 & 14.4 & 32.5 & 5.57 & $1.5\ts10^{-6}$ & 1620 & $4.2\ts10^{-7}$ \\
&400-3& 40000 & 15.0 & 39.2 & 5.71 & $2.7\ts10^{-6}$ & 1320 & $9.3\ts10^{-7}$ \\
&425-3& 42500 & 15.6 & 47.4 & 5.85 & $4.5\ts10^{-6}$ & 1200 & $1.5\ts10^{-6}$ \\
\hline Supergiants
&300-1& 30000 & 22.4 & 28.8 & 5.56 & $9.9\ts10^{-7}$ & 1110 & $4.5\ts10^{-7}$ \\
&325-1& 32500 & 21.4 & 34.0 & 5.66 & $1.7\ts10^{-6}$ & 1140 & $5.7\ts10^{-7}$ \\
&350-1& 35000 & 20.5 & 40.4 & 5.75 & $2.6\ts10^{-6}$ & 1350 & $6.9\ts10^{-7}$ \\
&375-1& 37500 & 19.8 & 48.3 & 5.84 & $3.7\ts10^{-6}$ & 1260 & $1.2\ts10^{-6}$ \\
&400-1& 40000 & 19.1 & 58.1 & 5.92 & $4.7\ts10^{-6}$ & 1460 & $1.4\ts10^{-6}$ \\
&425-1& 42500 & 18.5 & 70.3 & 6.00 & $5.6\ts10^{-6}$ & 1350 & $1.8\ts10^{-6}$ \\
\hline
\end{tabular}
\end{table*}

Deep in the stellar atmosphere the specific intensity of radiation is nearly
isotropic. Nevertheless, even the small deviations from isotropy are important,
because they drive the transport of radiation energy. Therefore, the radiative
transport can be described by the diffusion equation. This equation determines
the temperature gradient which is necessary to transfer the given flux through
the optically thick parts of the stellar atmosphere. Therefore, deep in the
stellar atmosphere we use the so-called differential form of the radiative
equilibrium equation to estimate the temperature \citep[similarly as][]{kubii}
\begin{equation}
\label{integ}
H(r)=\frac{\sigma T_\text{eff}^4}{4\pi}\frac{R^2_*}{r^2}=
-\int_0^\infty\frac{1}{q_\nu\chi_\nu}\frac{\de\zav{q_\nu f_\nu J_\nu}}
{\de r}
\de\nu,
\end{equation}
where $q_\nu$ and $f_\nu$ are sphericity and Eddington factors \citep{auer} and
$H(r)$ is the frequency-integrated Eddington flux derived from the CMF solution.
The mean intensity $J_\nu$ is evaluated after \eqref{uziteJ}. The sphericity
factor is defined as $(1/q_\nu)(\de q_\nu/\de r) = (3-f_\nu^{-1})/r$. Equation
\eqref{integ} also sets the luminosity of the model with a fixed stellar
effective temperature $T_\text{eff}$ and radius\footnote{We have not changed the
definition of the stellar radius $R_*$ from our previous models, that is, $R_*$
corresponds to the innermost grid point of the models. This brings some
difference with respect to commonly used definitions of $R_*$. However, this
difference is large only for stars with extended photospheres, which are not
studied here.} $R_*$. The equation $H(r)={\sigma
T_\text{eff}^4R^2_*/}{(4\pi}{r^2)}$ is solved using Newton-Raphson procedure
together with remaining hydrodynamical equations. The derivatives of the CMF
flux $H(r)$ with respect to temperature for the Newton-Raphson iterations are
calculated from the integral on the right hands side of Eq.~\eqref{integ}. The
required derivatives of $J_\nu$ are calculated from the momentum form of the
continuum radiative transfer equation \citep[Eq.~(4)]{kubii} with derivatives of
the level populations with respect to the flow variables obtained from kinetic
equilibrium equations.

In the upper layers of the atmosphere, a significant part of the radiation
escapes and the the radiative transport is no more a diffusion process. The
model temperature in these layers is determined as a temperature, for which the
energy absorbed by the material is equal to the emitted energy and at which
the flux is therefore conserved,
\begin{equation}
\label{diferencial}
\int_0^\infty\zav{\eta_\nu-\chi_\nu J_\nu^\text{cont}}\,\de\nu=
\frac{1}{r^2}\frac{\de\!\zav{r^2 H (r) }}{\de r}= 0.
\end{equation}
Because the CMF flux $H(r)$ is a crucial quantity for the calculation of the
radiative force, we use equation $ r^{-2}\, \de \! \zav{r^2 H (r) } \! /{\de
r}=0$ to derive the temperature within the Newton-Raphson iterations of
hydrodynamical structure. However, the derivatives of $ r^{-2}\, \de \! \zav{r^2
H (r) } \! /{\de r}$ with respect to the flow variables are calculated using
left hand side of Eq.~\eqref{diferencial} written for continuum only. This
approximation affects only the values of the derivatives, that is the convergence
rate, but not the converged values, which are set by the CMF flux. This
procedure differs from that used in \cite{kubii}.

Concerning the radiative processes, only bound-free and free-free processes
directly affect the temperature. This has not significant consequences in the
photosphere, but has profound implications in the wind, where the bound-bound
processes dominate. Therefore, we
use
the electron thermal balance method, which
includes only processes that directly influence the temperature to calculate the
radiative heating term \citep{kpp}. This term is included in energy equation
\citep[Appendix~C]{nltei}.

The division between the individual parts of the model, where we used different
equation for the temperature, is fixed and given as an input parameter for each
model. A proper choice of these division points depends on resulting model
atmosphere, which is not known at the beginning of the calculation.
Consequently, the initial estimation of these parameters is not in all cases ideal
and does not always lead to the conservation of the radiative flux. In such
cases we recalculate the models with improved guess of the parameters and with
the inconsistent model as an initial estimate of the atmospheric structure. The
total CMF radiation flux is conserved with relative accuracy of about $10^{-5}$
and better in the regions where we use the differential form of the radiative
transfer equation \eqref{integ}, with relative accuracy of $10^{-4}-10^{-3}$ in
the region in which we used the integral form Eq.~\eqref{diferencial}, and within
a few percent in the region in which we used the electron thermal balance method.

\begin{figure*}[tp]
\centering
\resizebox{0.310\hsize}{!}{\includegraphics{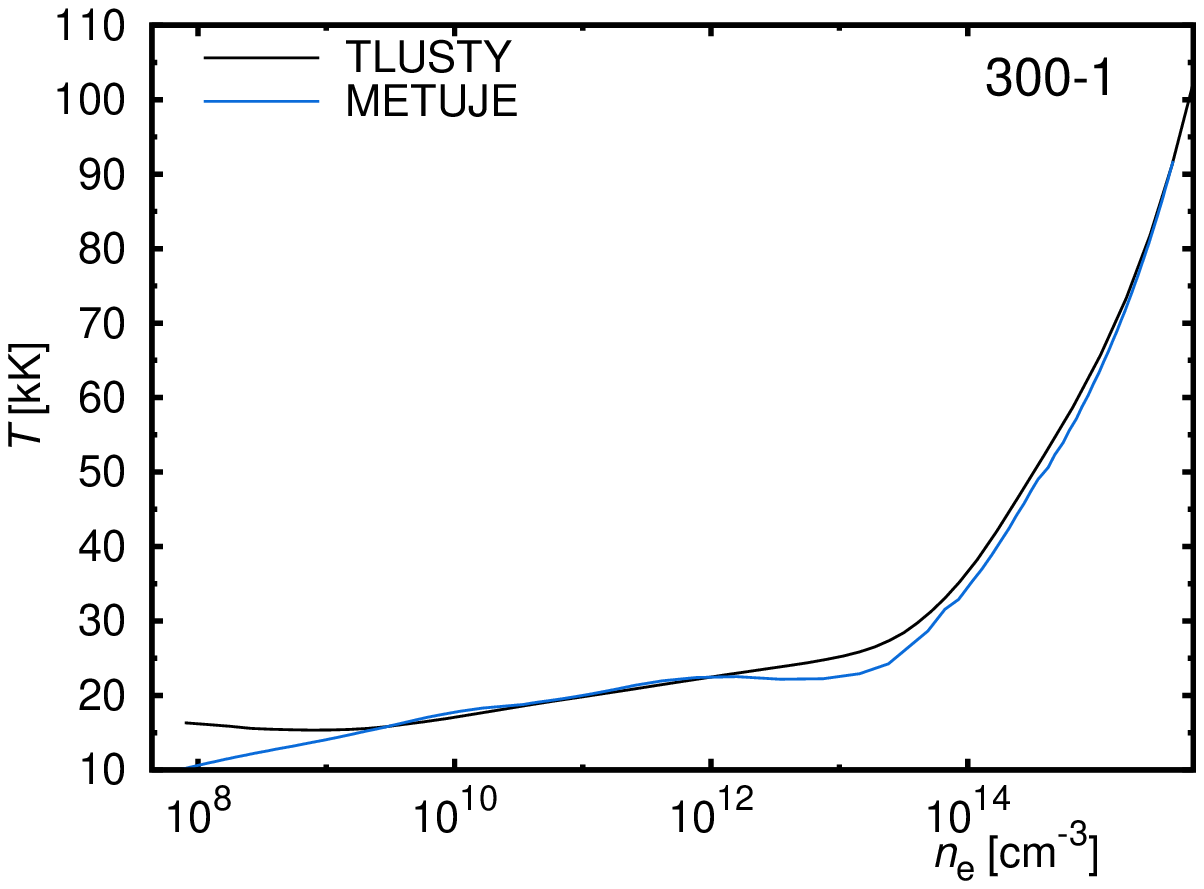}}
\resizebox{0.310\hsize}{!}{\includegraphics{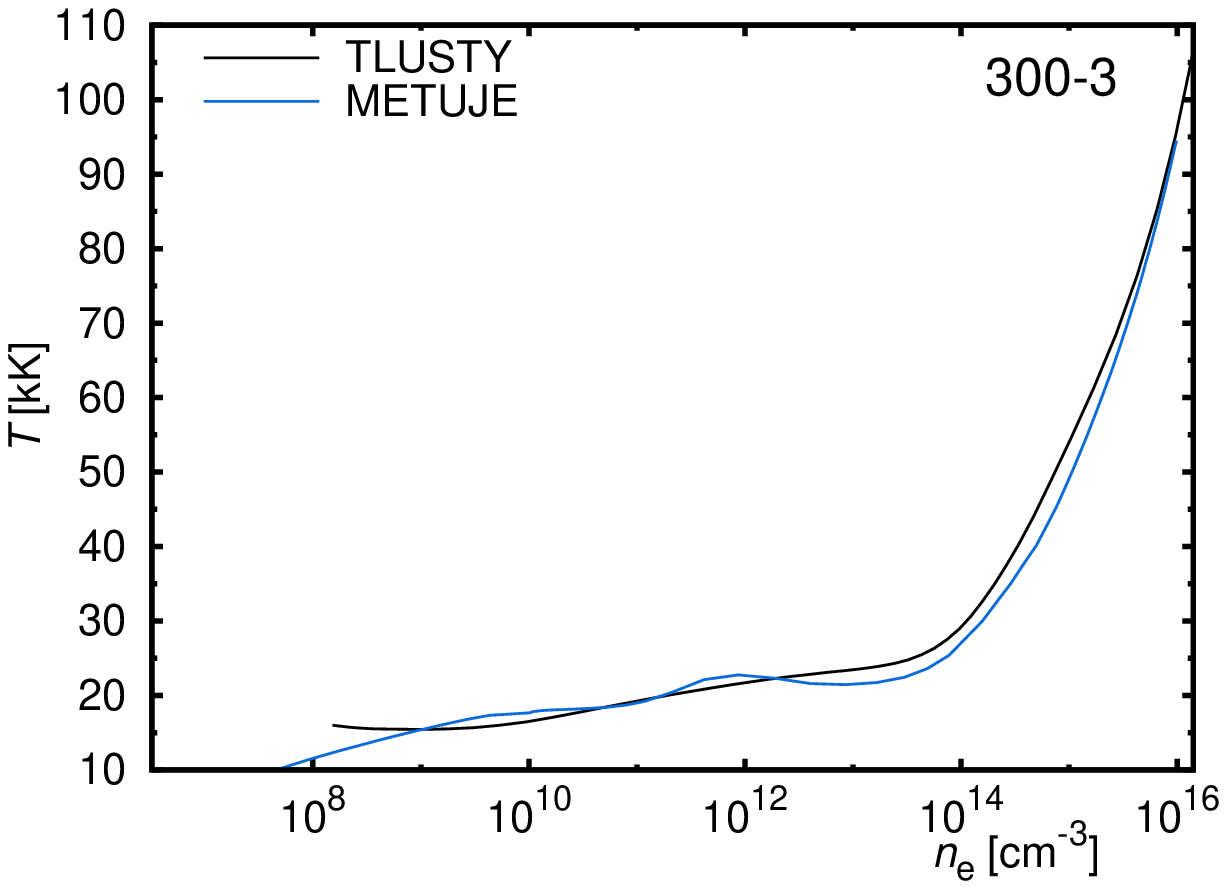}}
\resizebox{0.310\hsize}{!}{\includegraphics{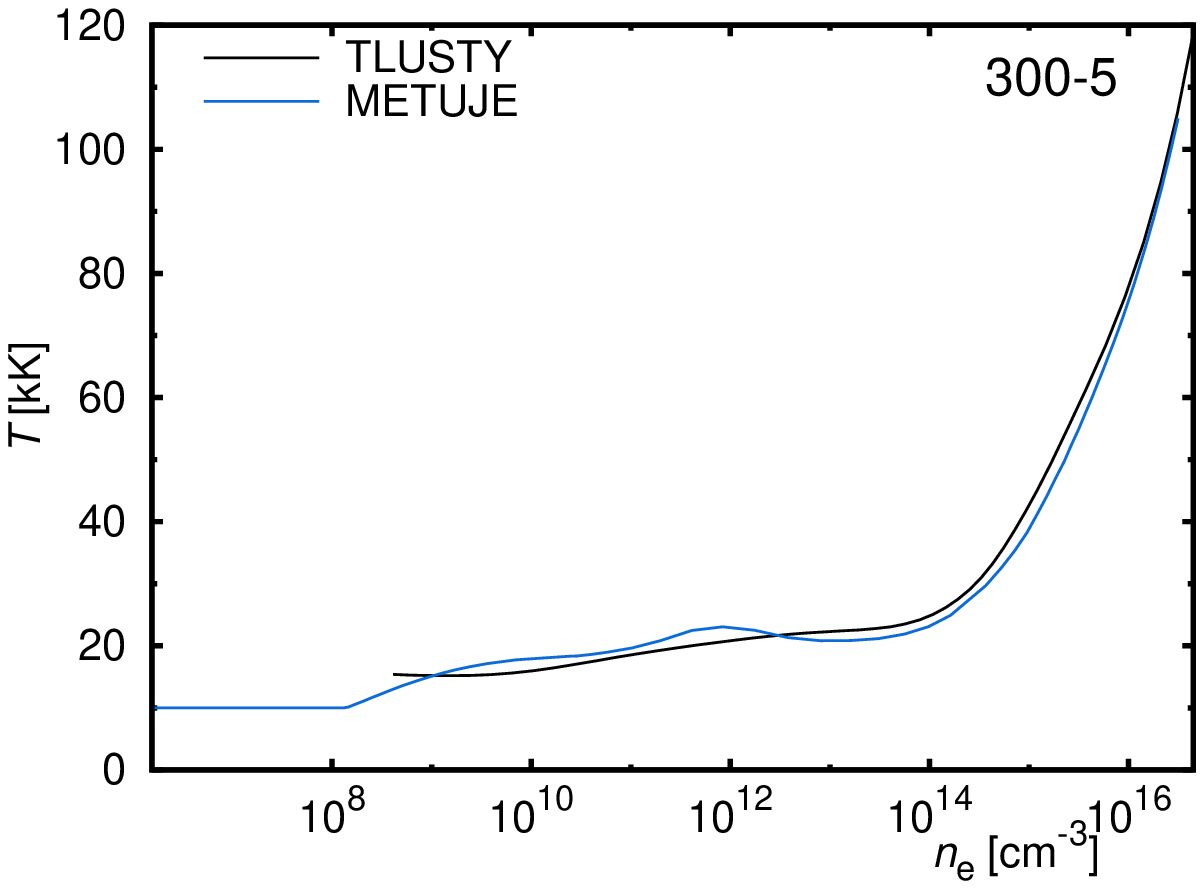}}\\
\resizebox{0.310\hsize}{!}{\includegraphics{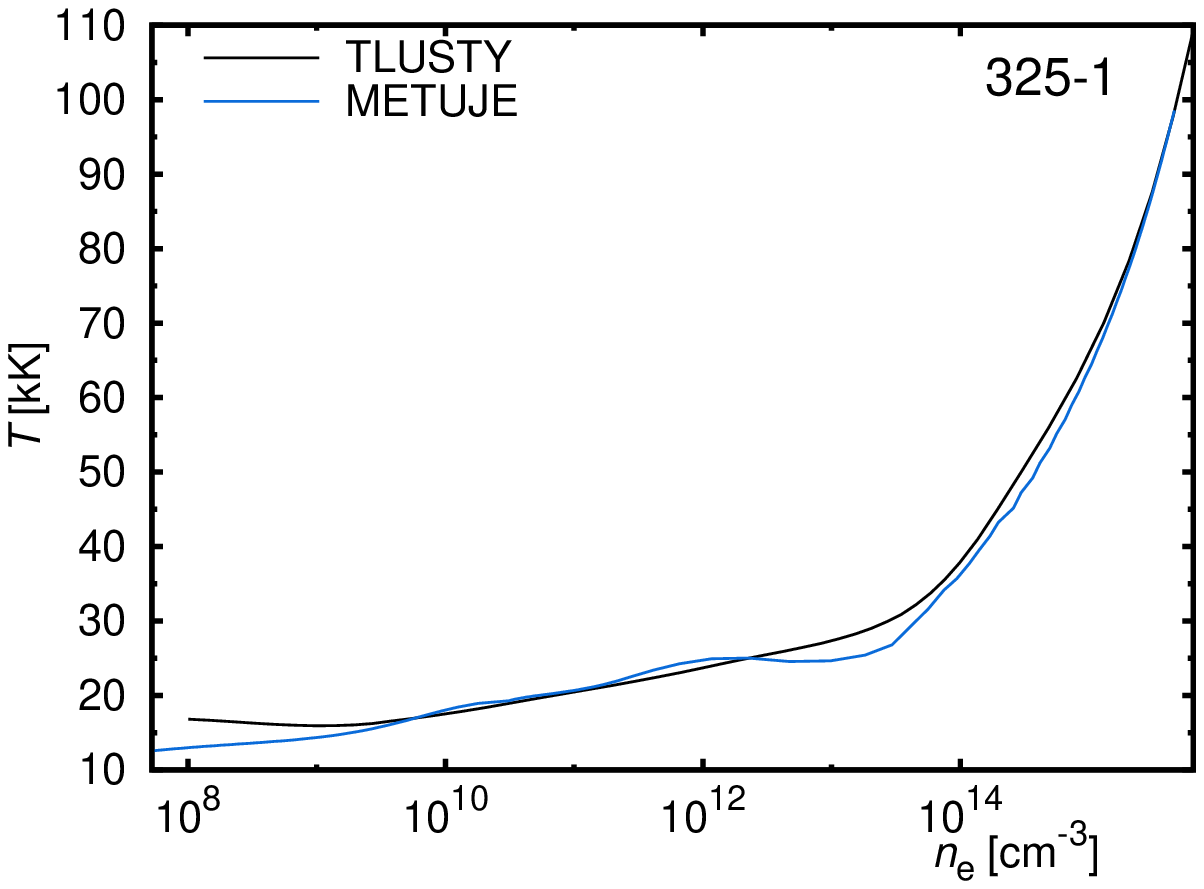}}
\resizebox{0.310\hsize}{!}{\includegraphics{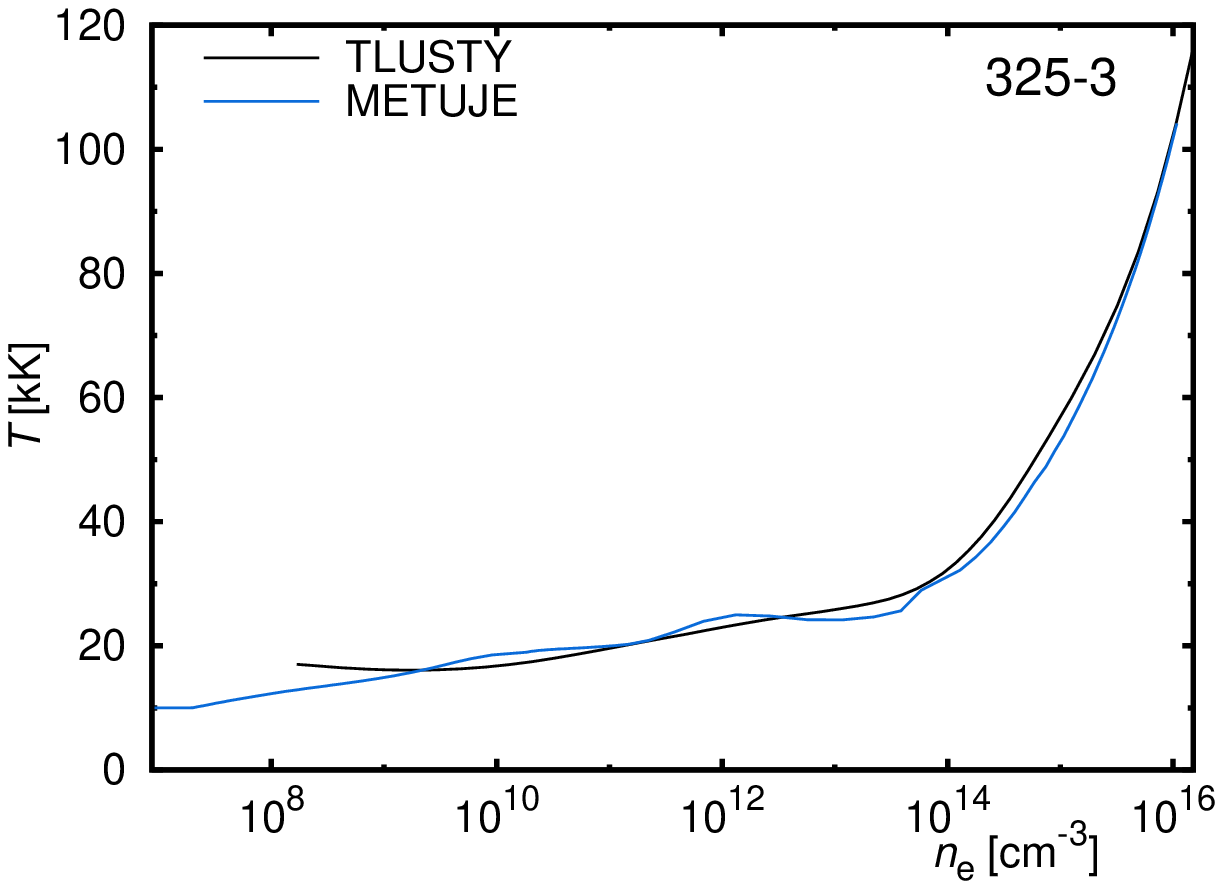}}
\resizebox{0.310\hsize}{!}{\includegraphics{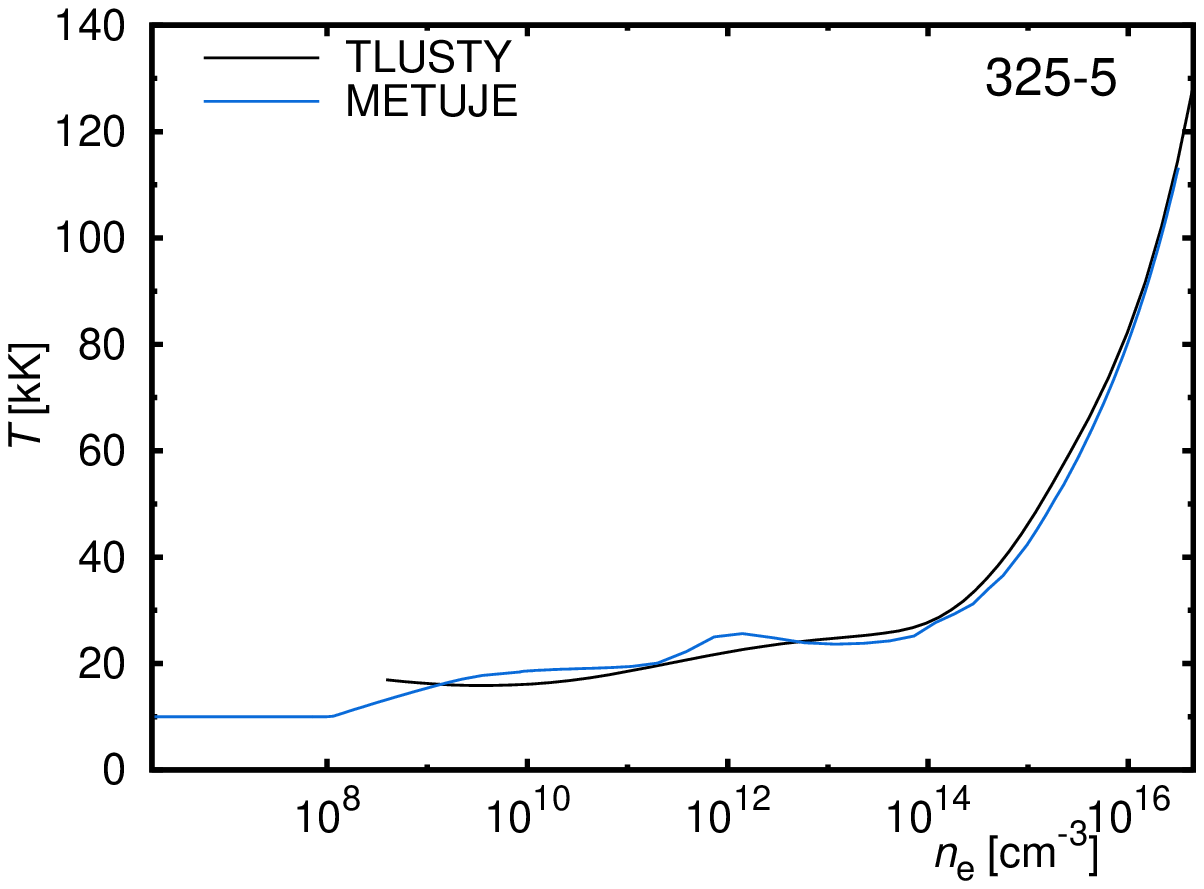}}\\
\resizebox{0.310\hsize}{!}{\includegraphics{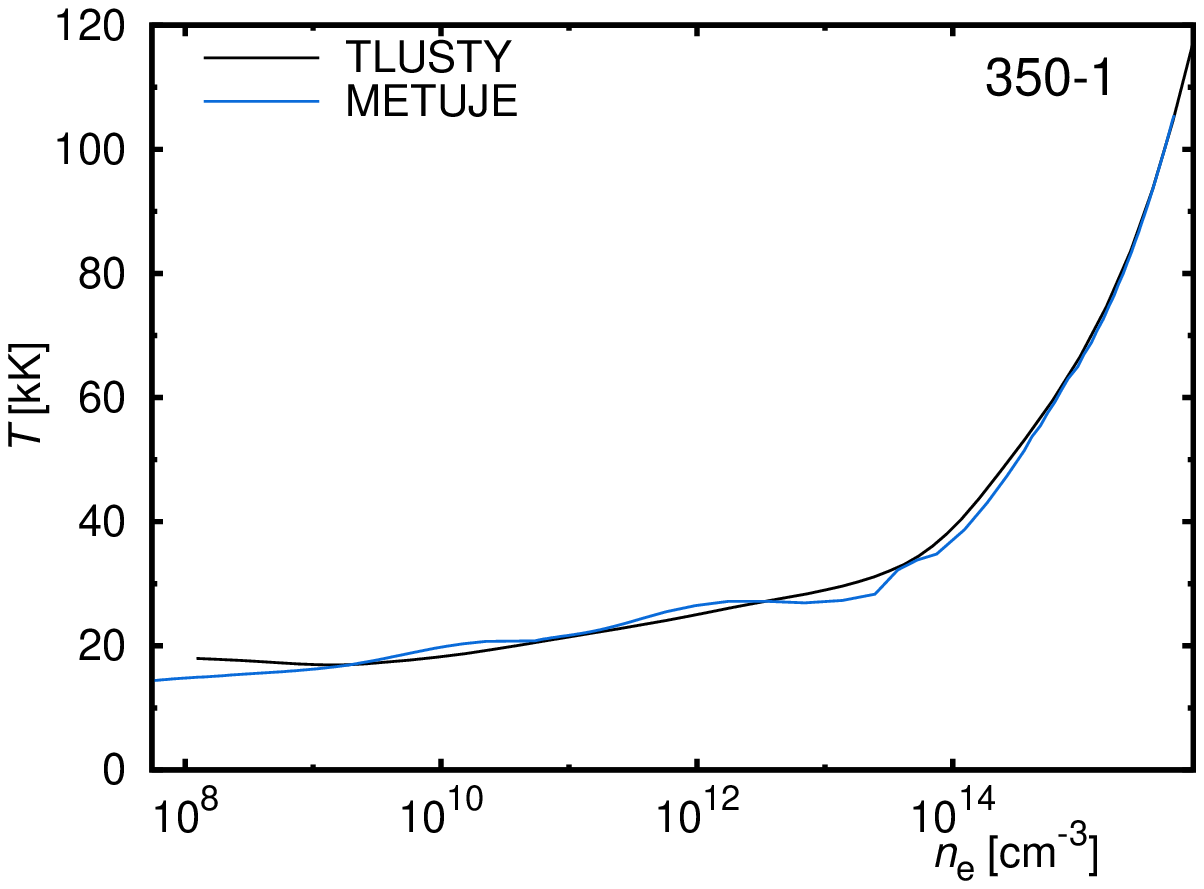}}
\resizebox{0.310\hsize}{!}{\includegraphics{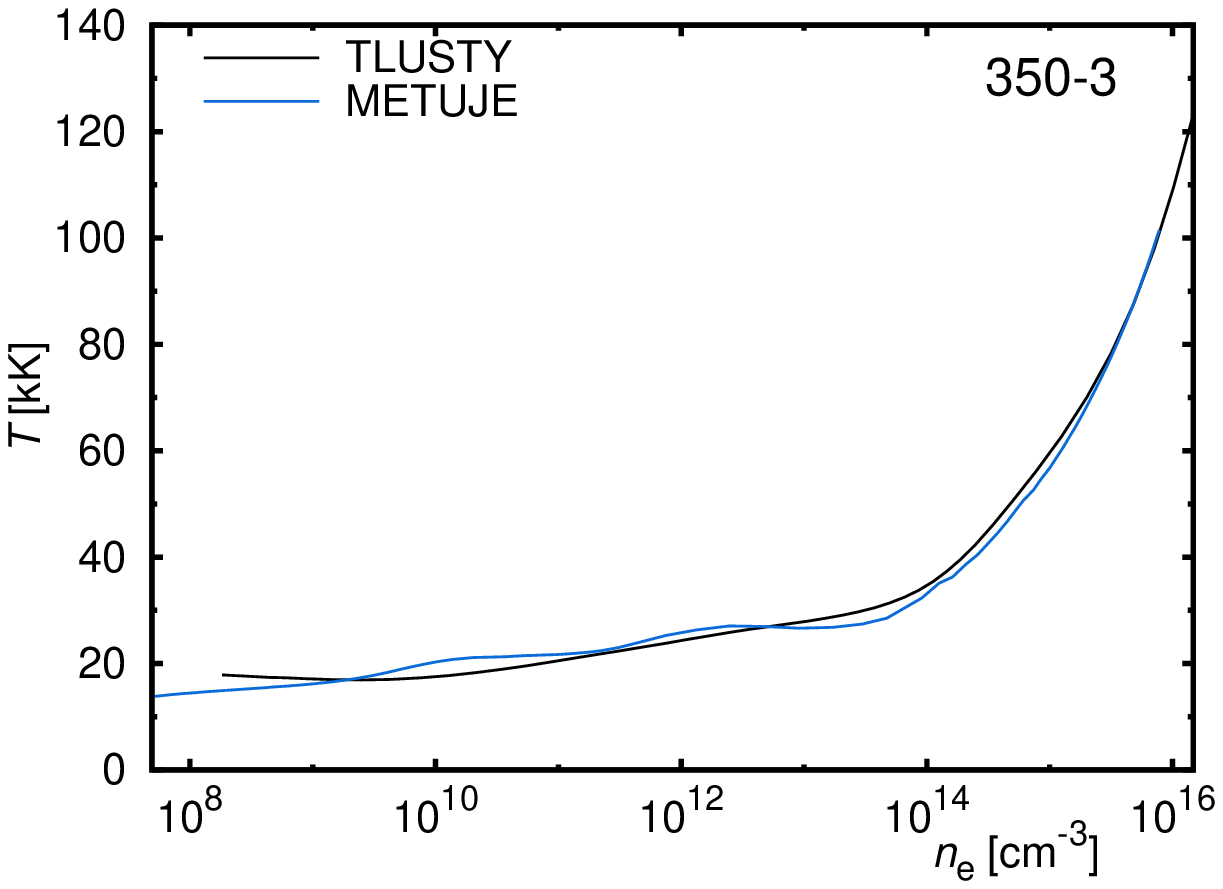}}
\resizebox{0.310\hsize}{!}{\includegraphics{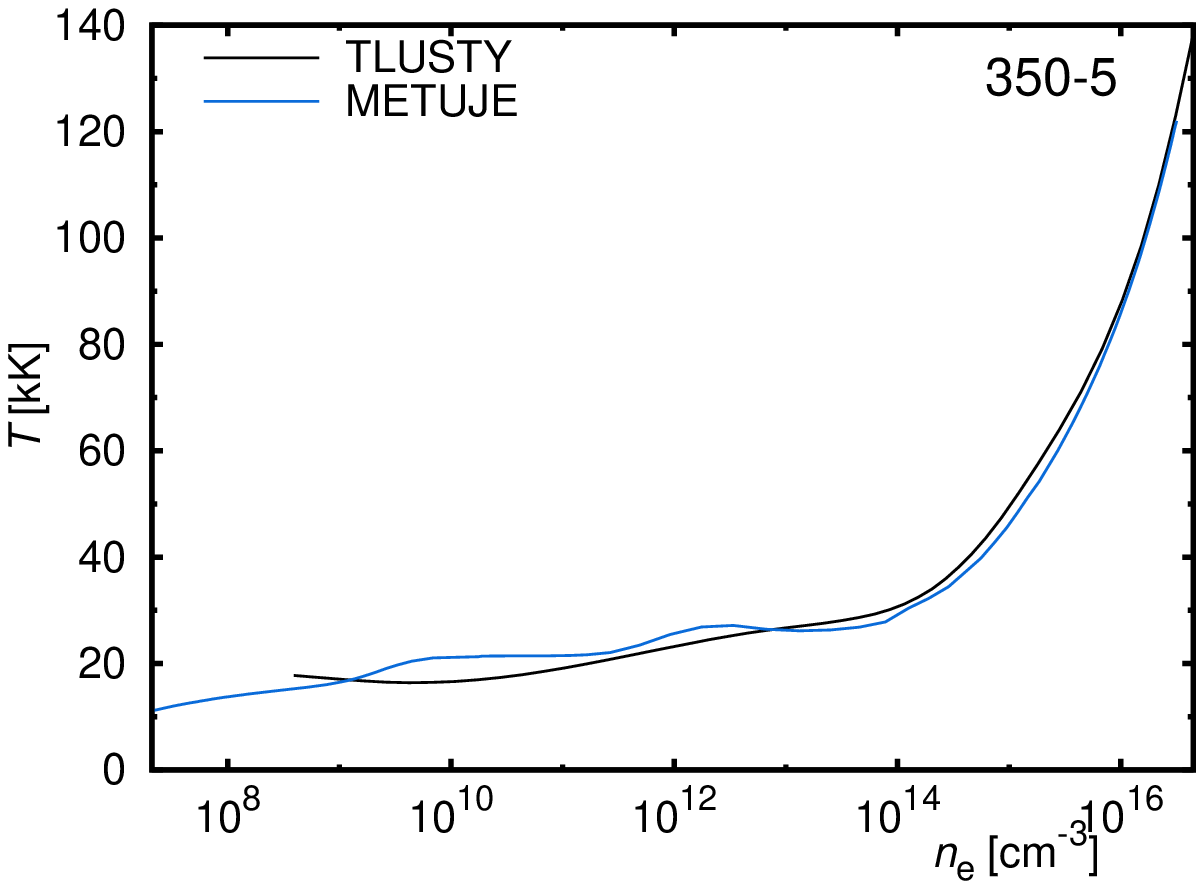}}\\
\resizebox{0.310\hsize}{!}{\includegraphics{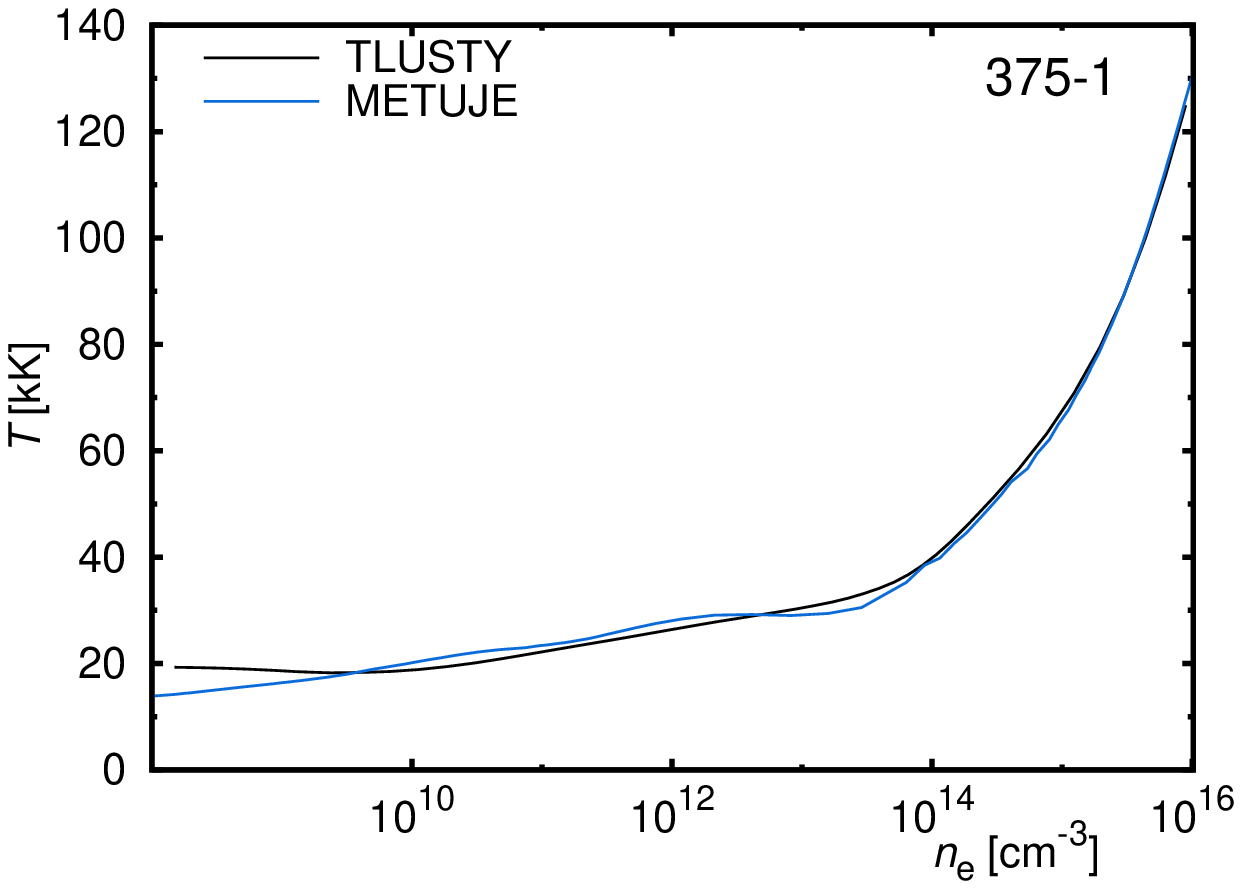}}
\resizebox{0.310\hsize}{!}{\includegraphics{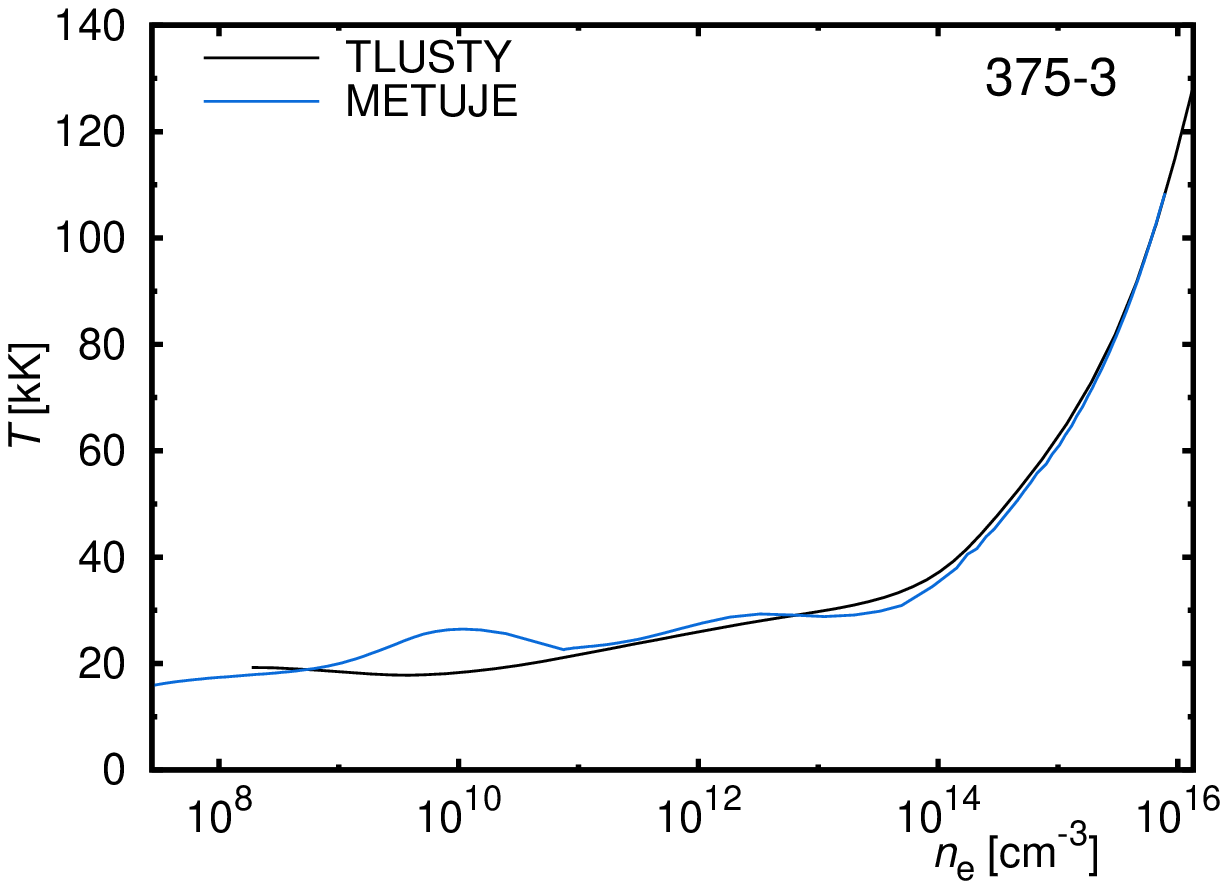}}
\resizebox{0.310\hsize}{!}{\includegraphics{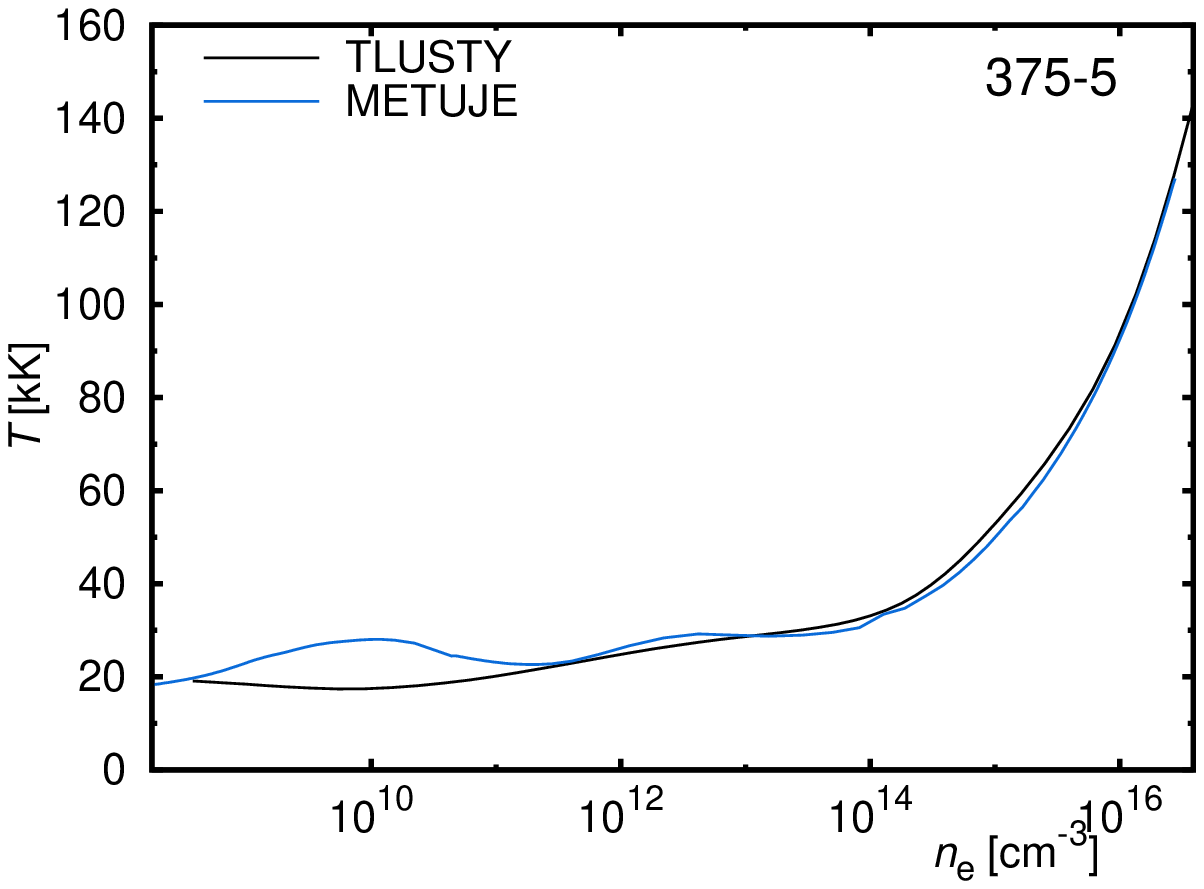}}\\
\resizebox{0.310\hsize}{!}{\includegraphics{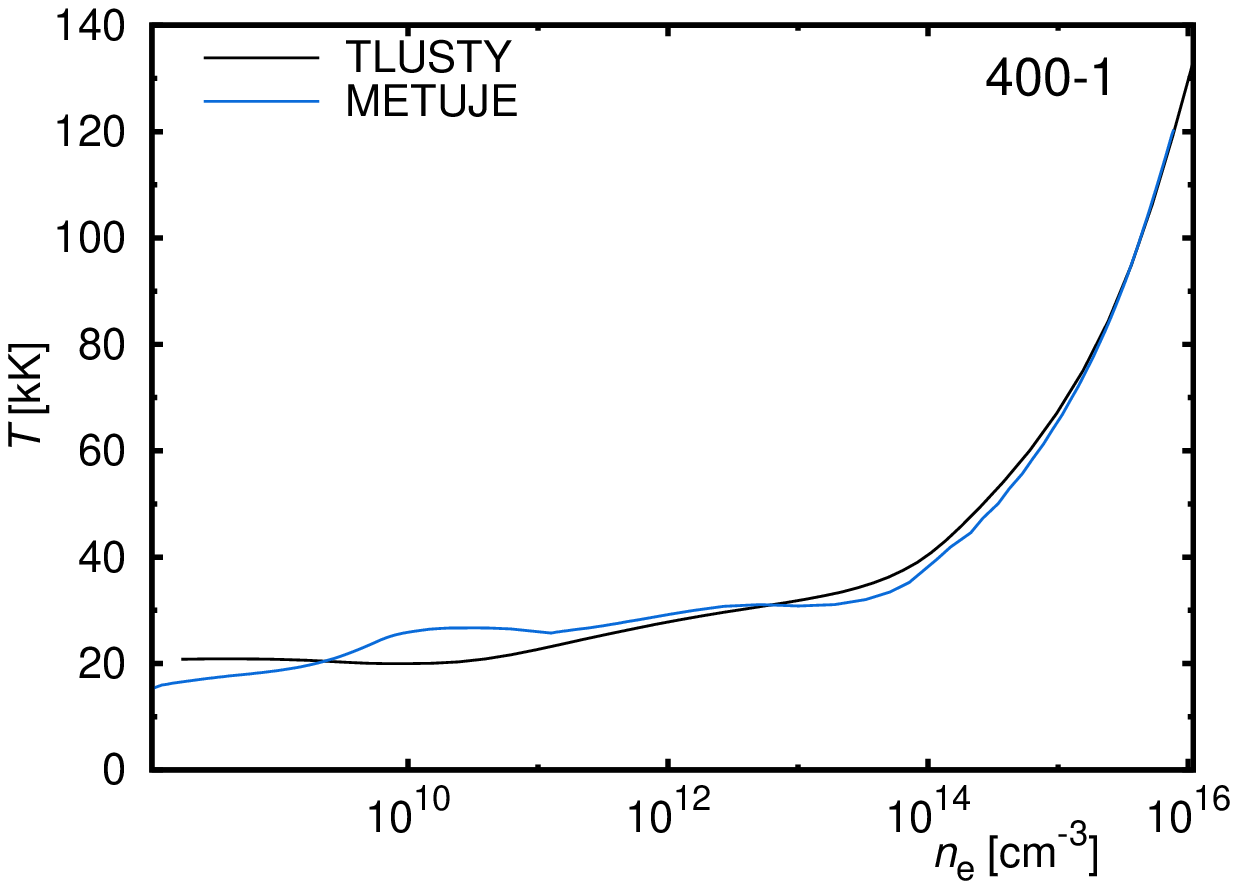}}
\resizebox{0.310\hsize}{!}{\includegraphics{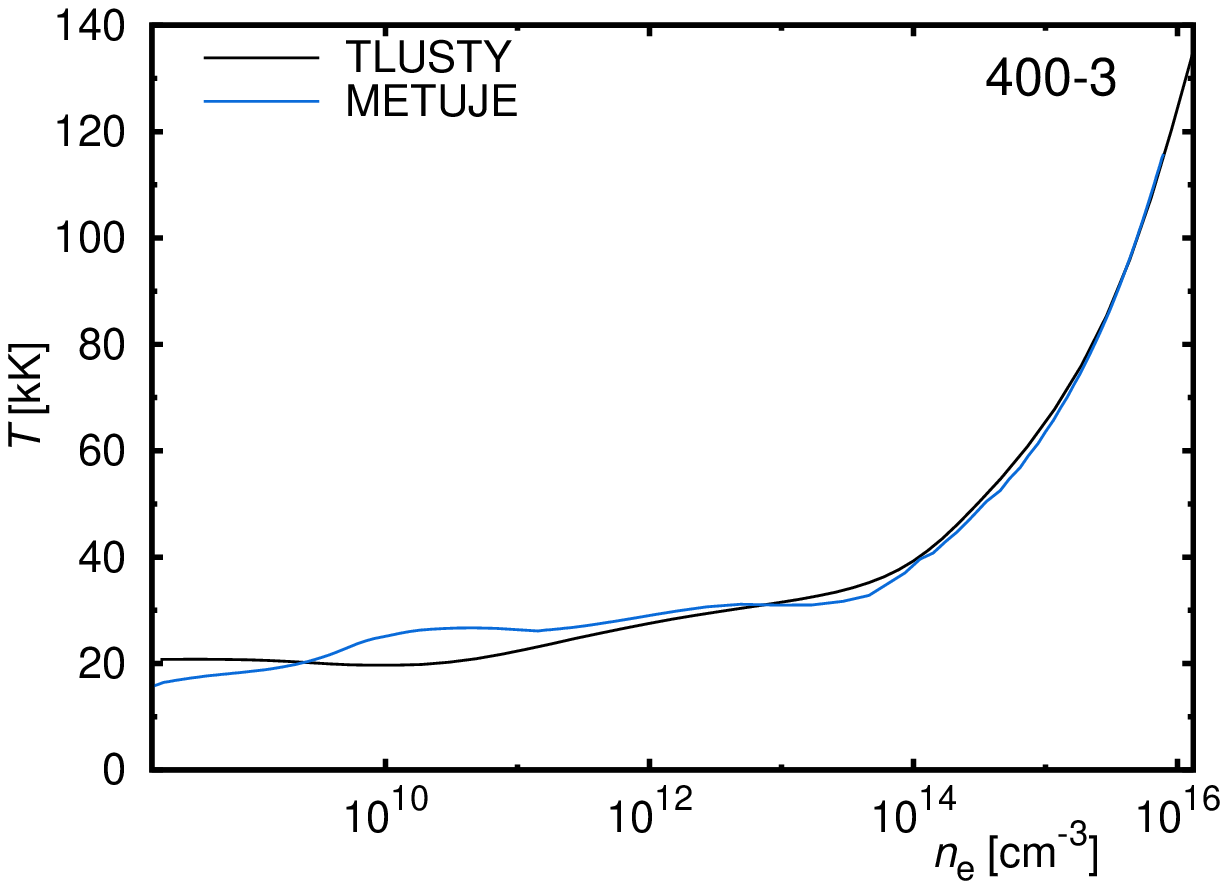}}
\resizebox{0.310\hsize}{!}{\includegraphics{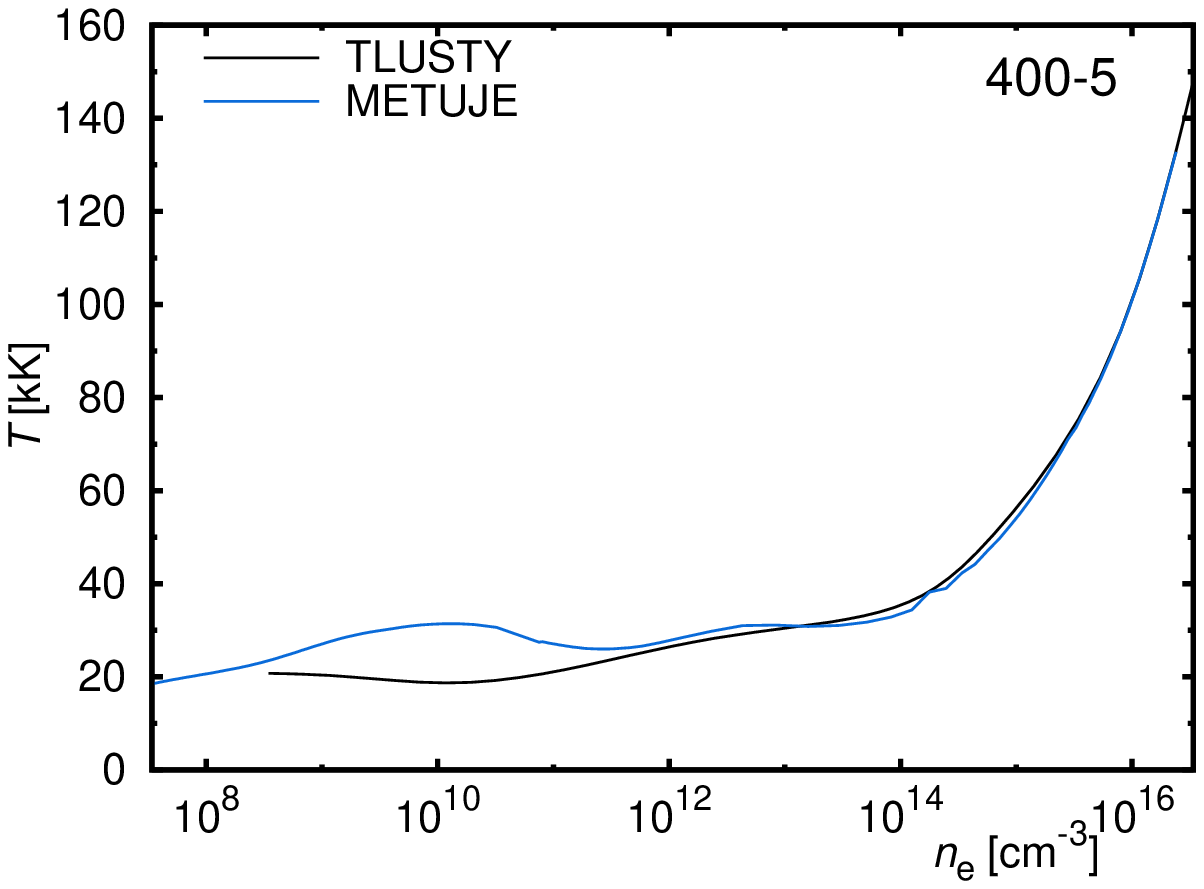}}\\
\resizebox{0.310\hsize}{!}{\includegraphics{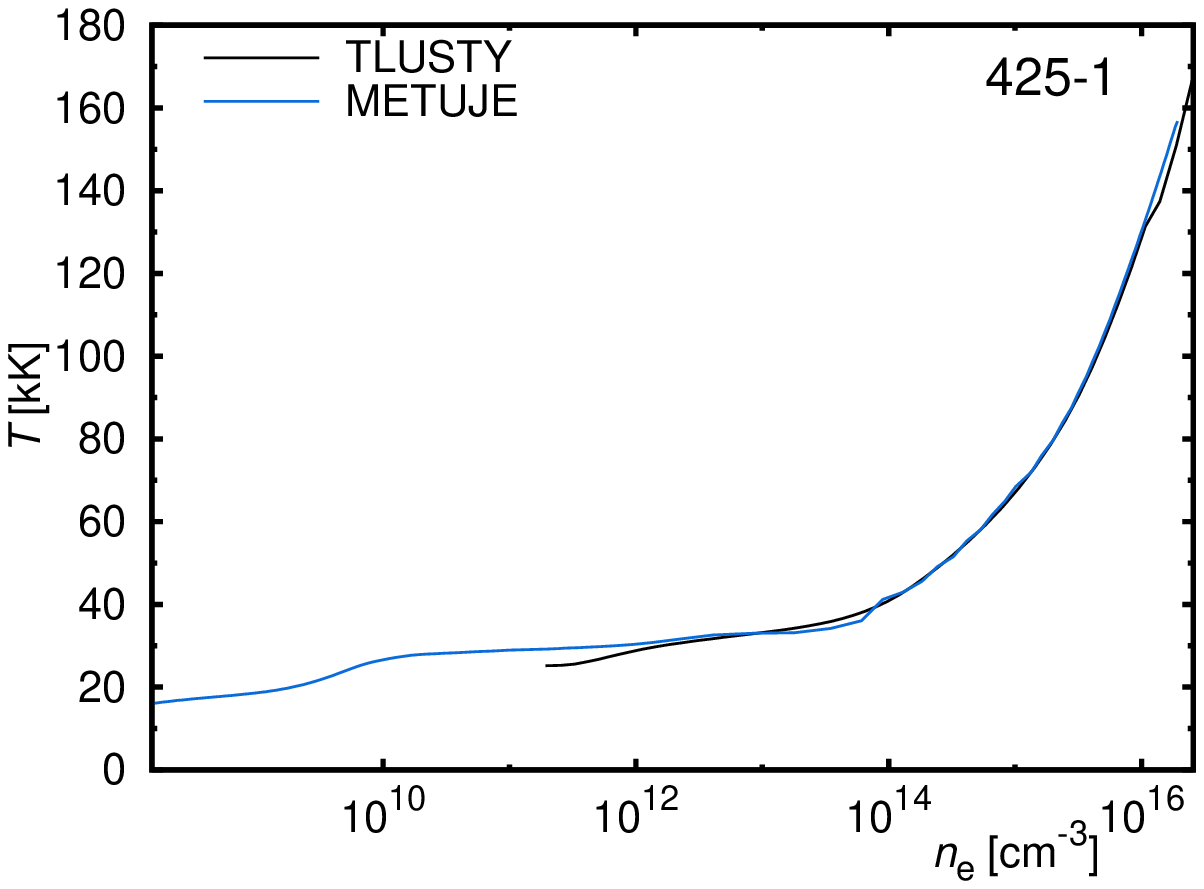}}
\resizebox{0.310\hsize}{!}{\includegraphics{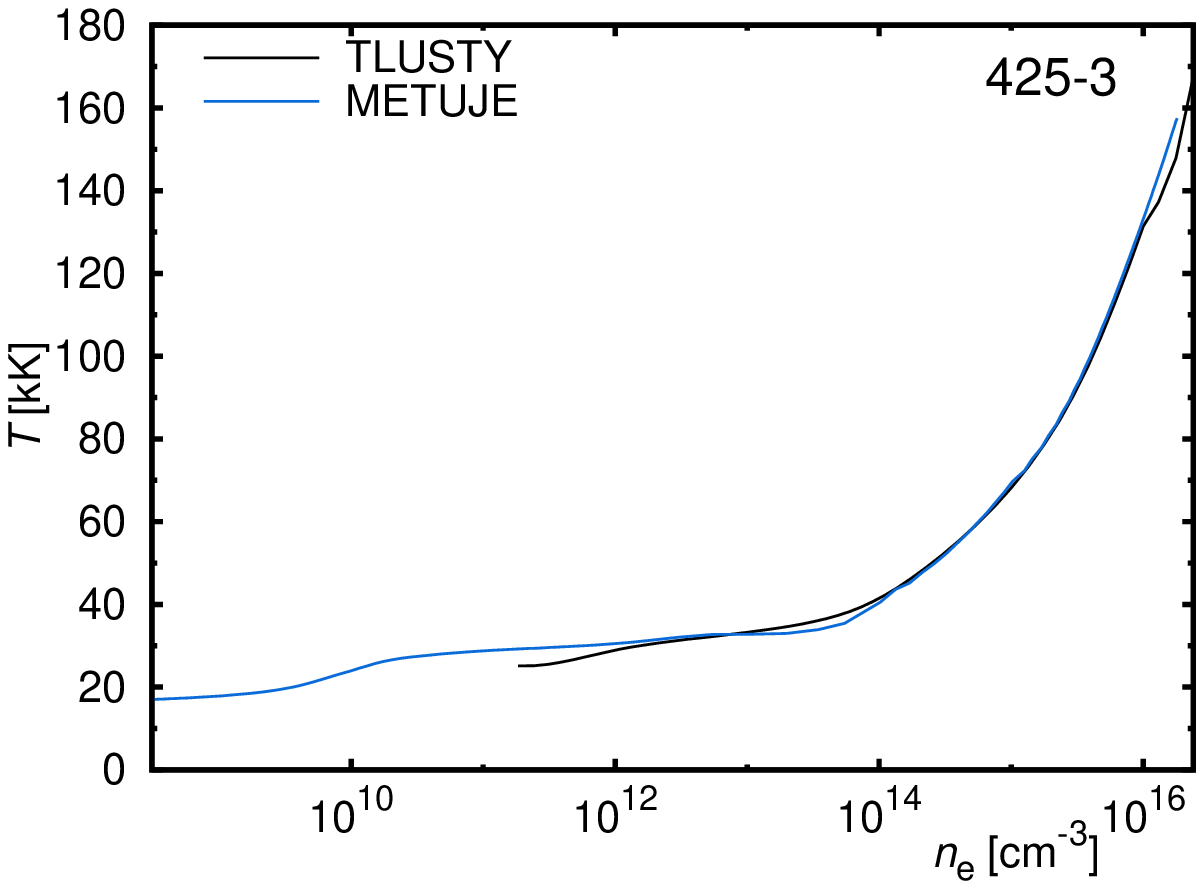}}
\resizebox{0.310\hsize}{!}{\includegraphics{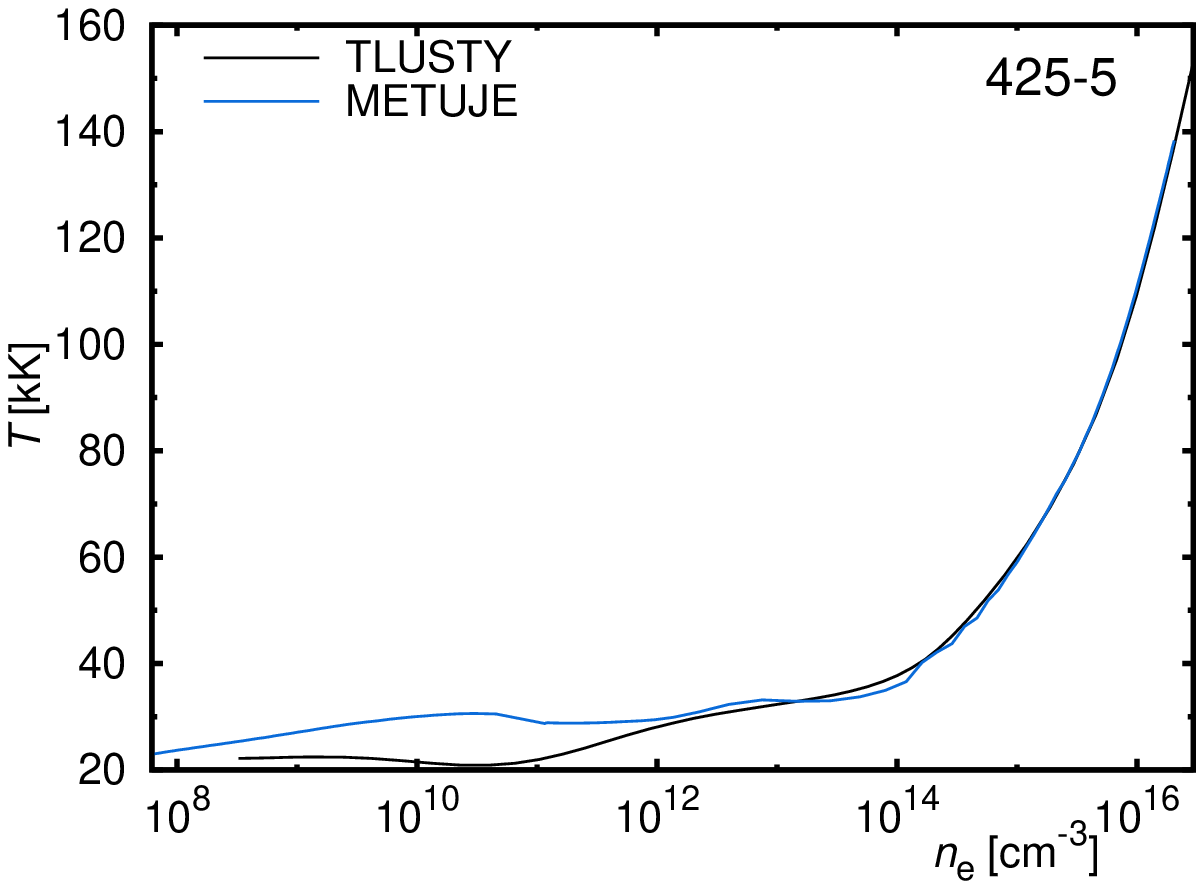}}
\caption{Comparison of the dependence of temperature on electron density in
TLUSTY and METUJE models. The graphs are plotted for individual model stars from
Table~\ref{ohvezpar} (denoted in the plots).}
\label{tepmetlu}
\end{figure*}

\begin{figure*}[tp]
\centering
\resizebox{0.310\hsize}{!}{\includegraphics{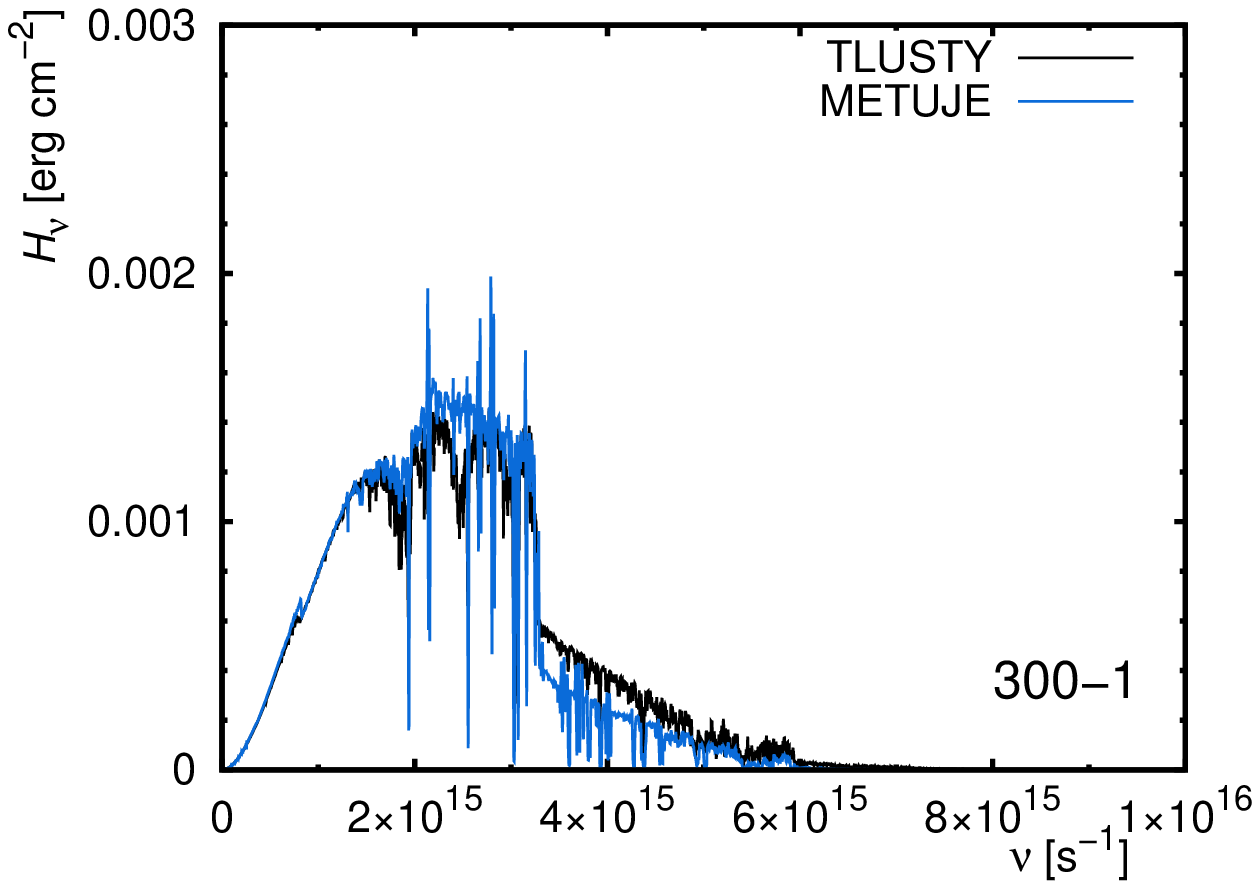}}
\resizebox{0.310\hsize}{!}{\includegraphics{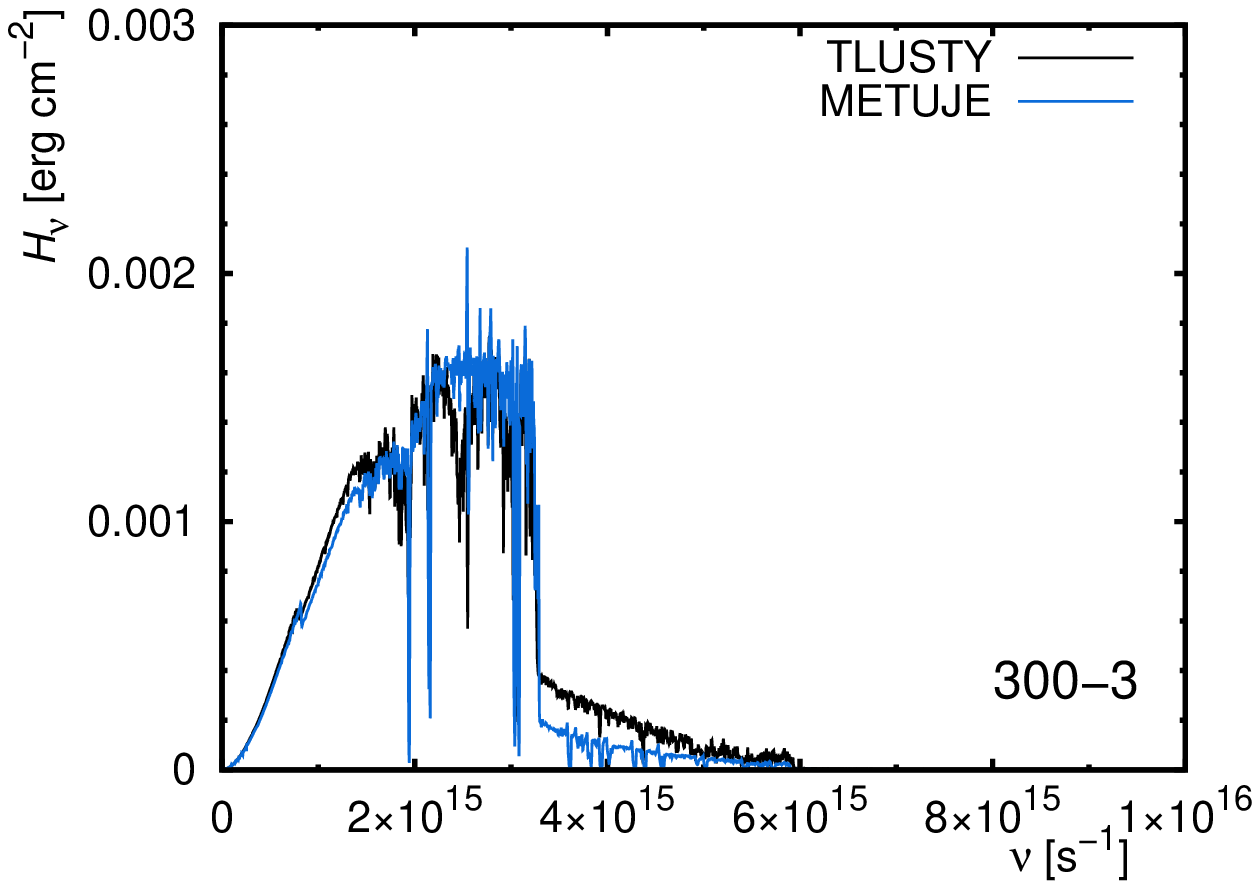}}
\resizebox{0.310\hsize}{!}{\includegraphics{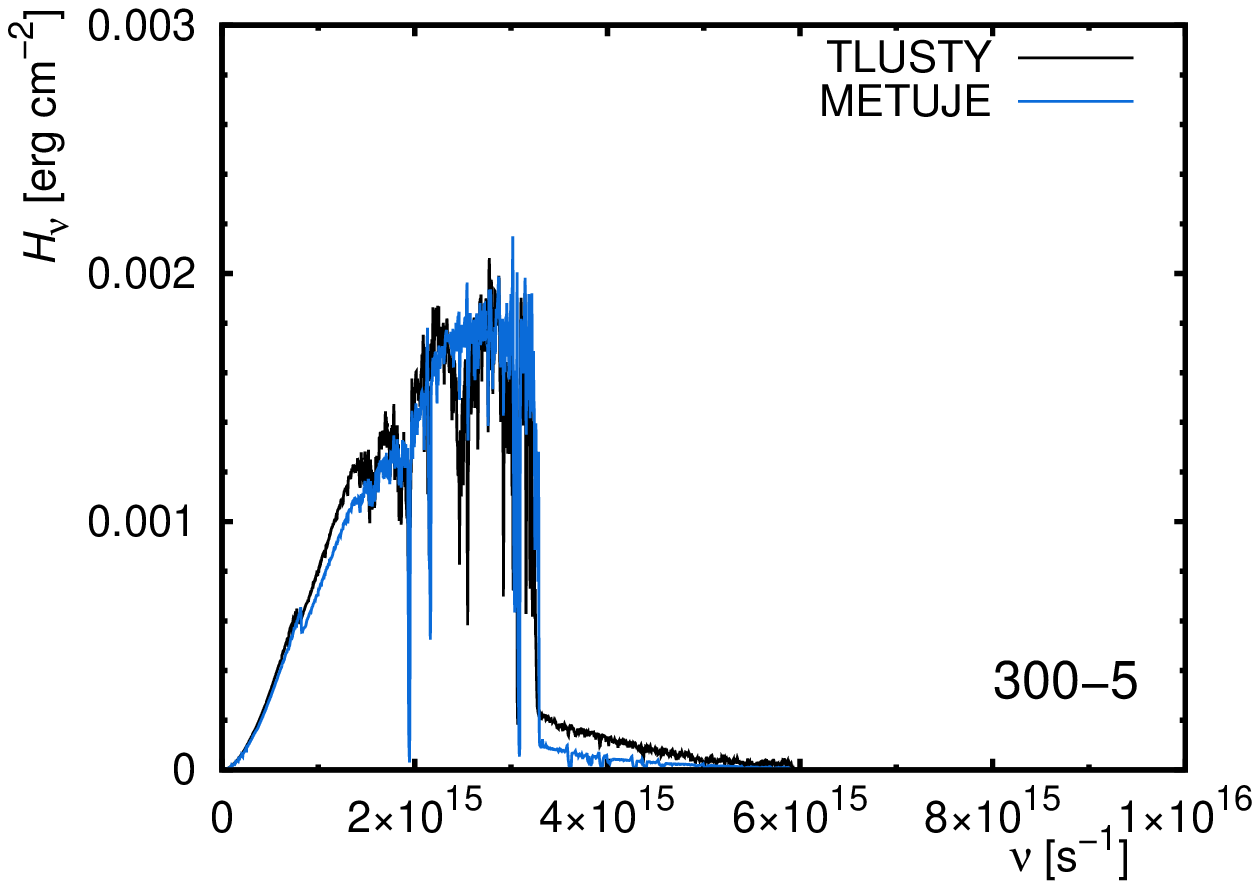}}\\
\resizebox{0.310\hsize}{!}{\includegraphics{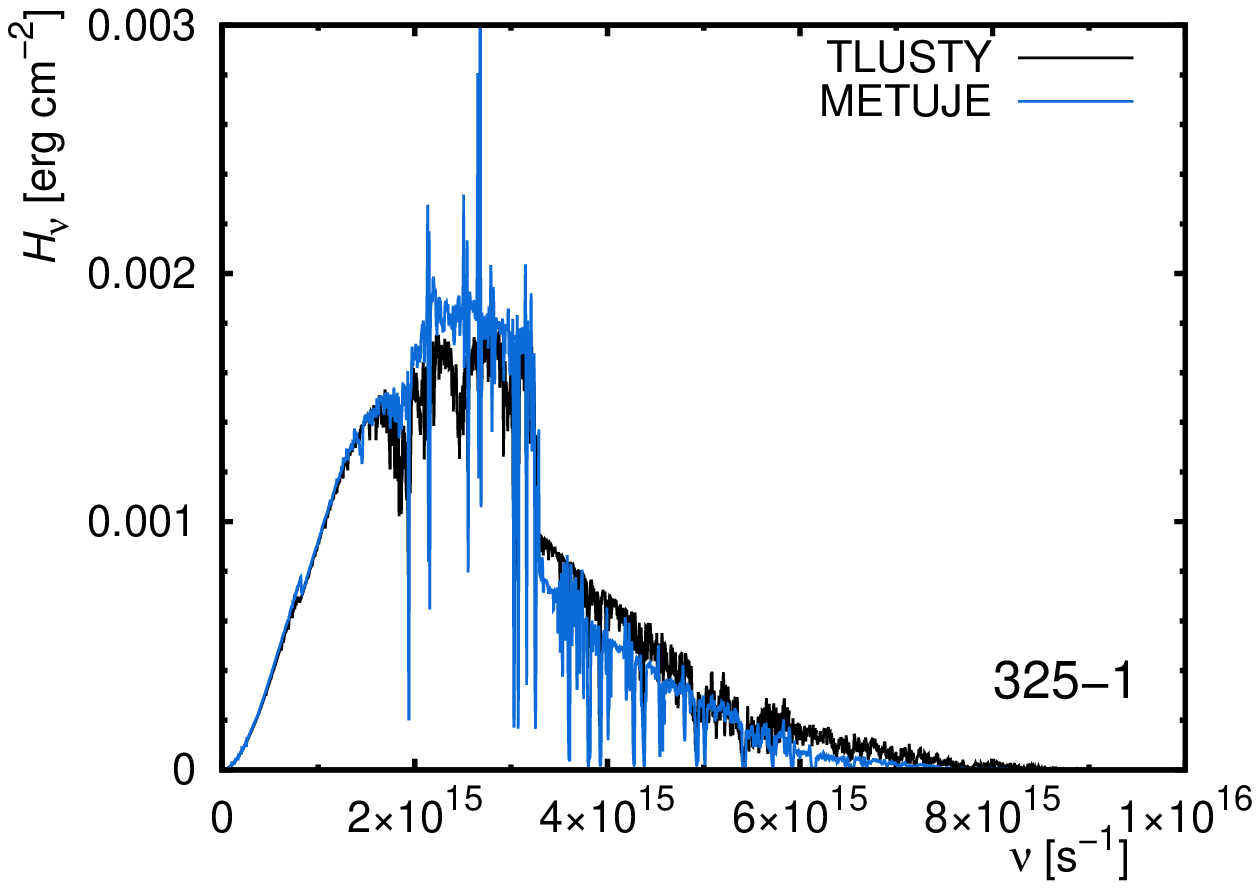}}
\resizebox{0.310\hsize}{!}{\includegraphics{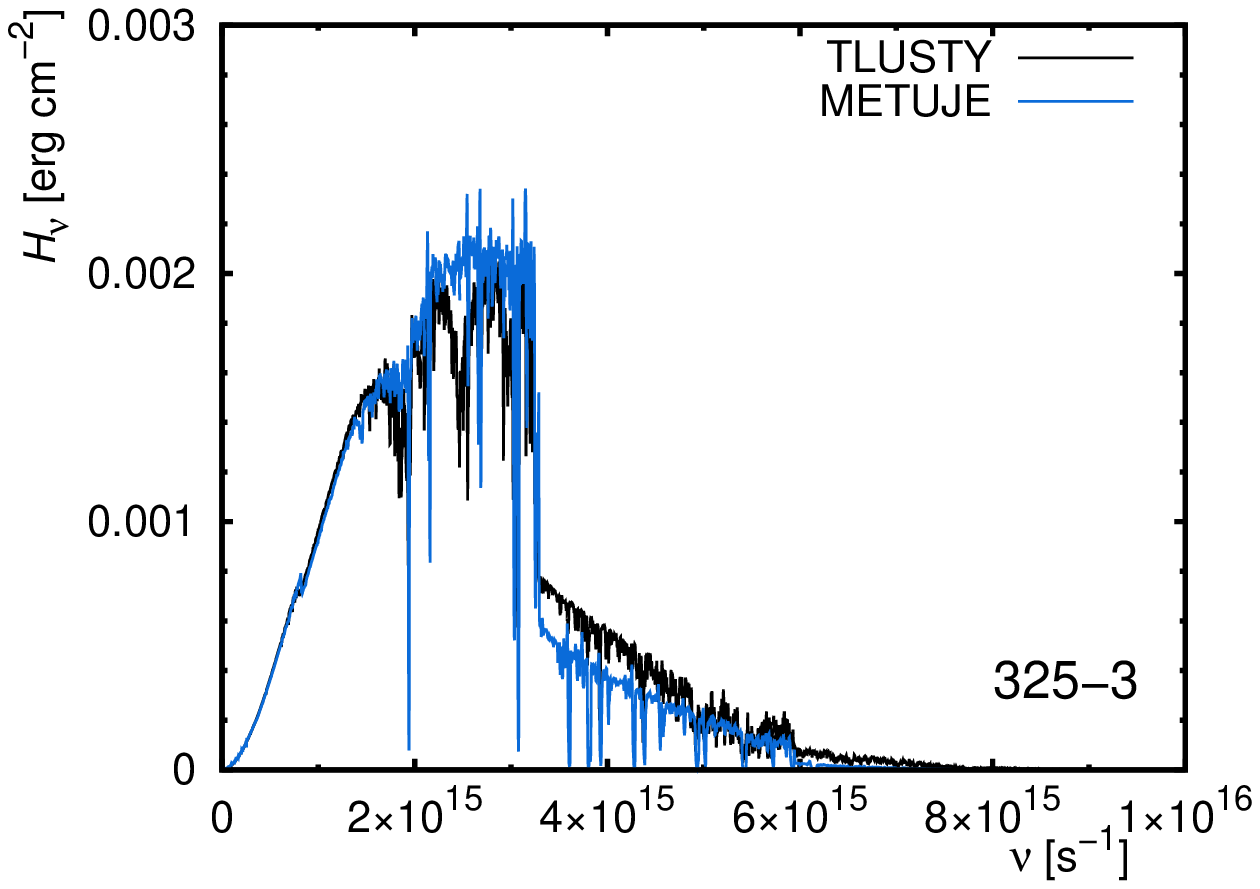}}
\resizebox{0.310\hsize}{!}{\includegraphics{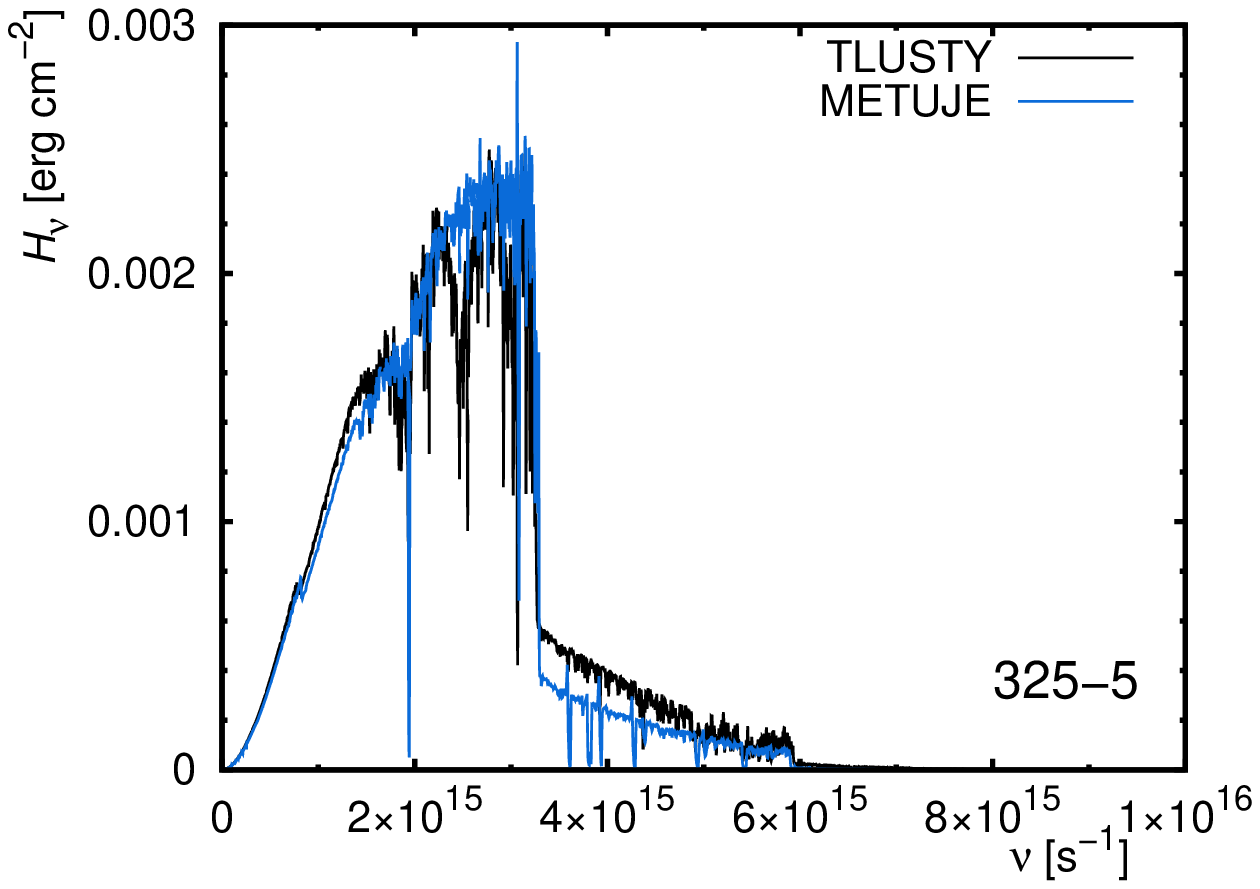}}\\
\resizebox{0.310\hsize}{!}{\includegraphics{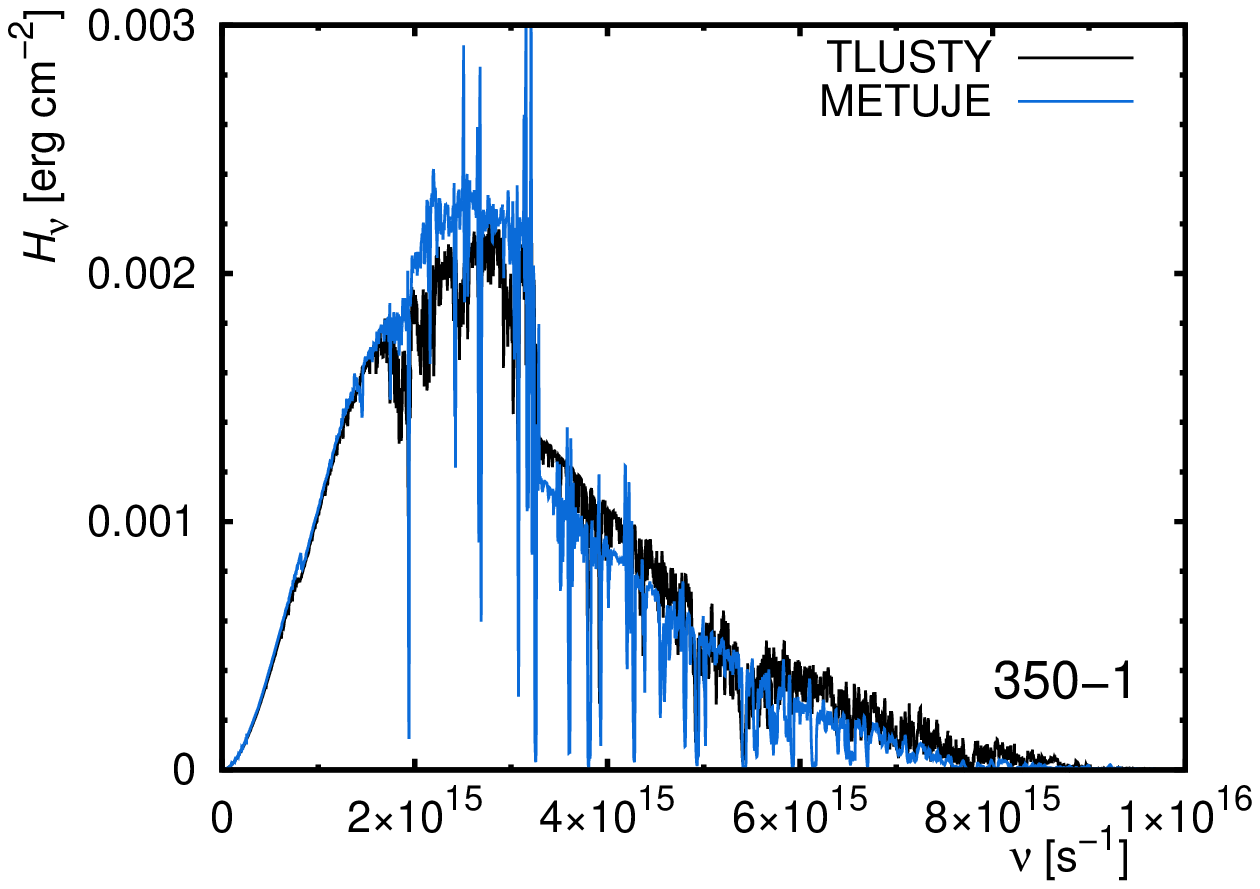}}
\resizebox{0.310\hsize}{!}{\includegraphics{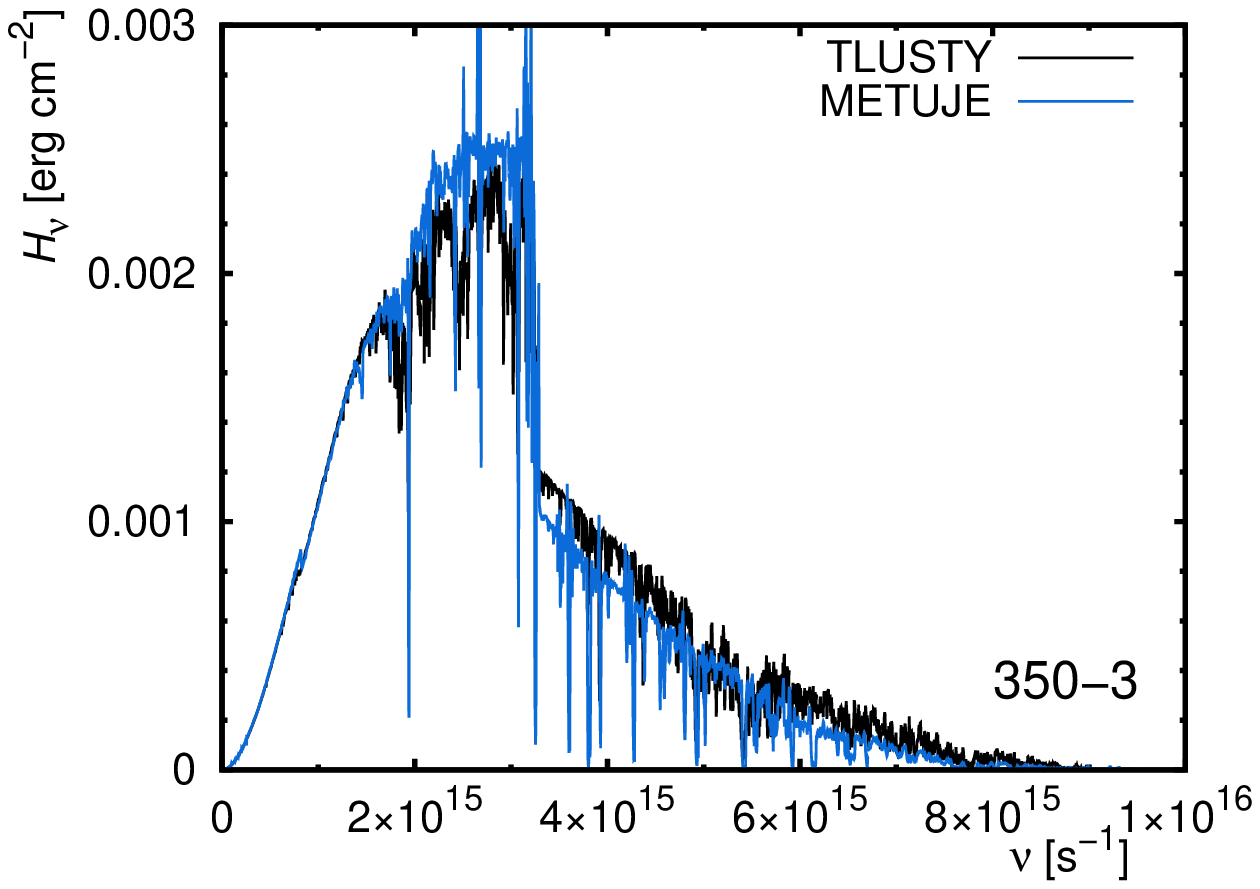}}
\resizebox{0.310\hsize}{!}{\includegraphics{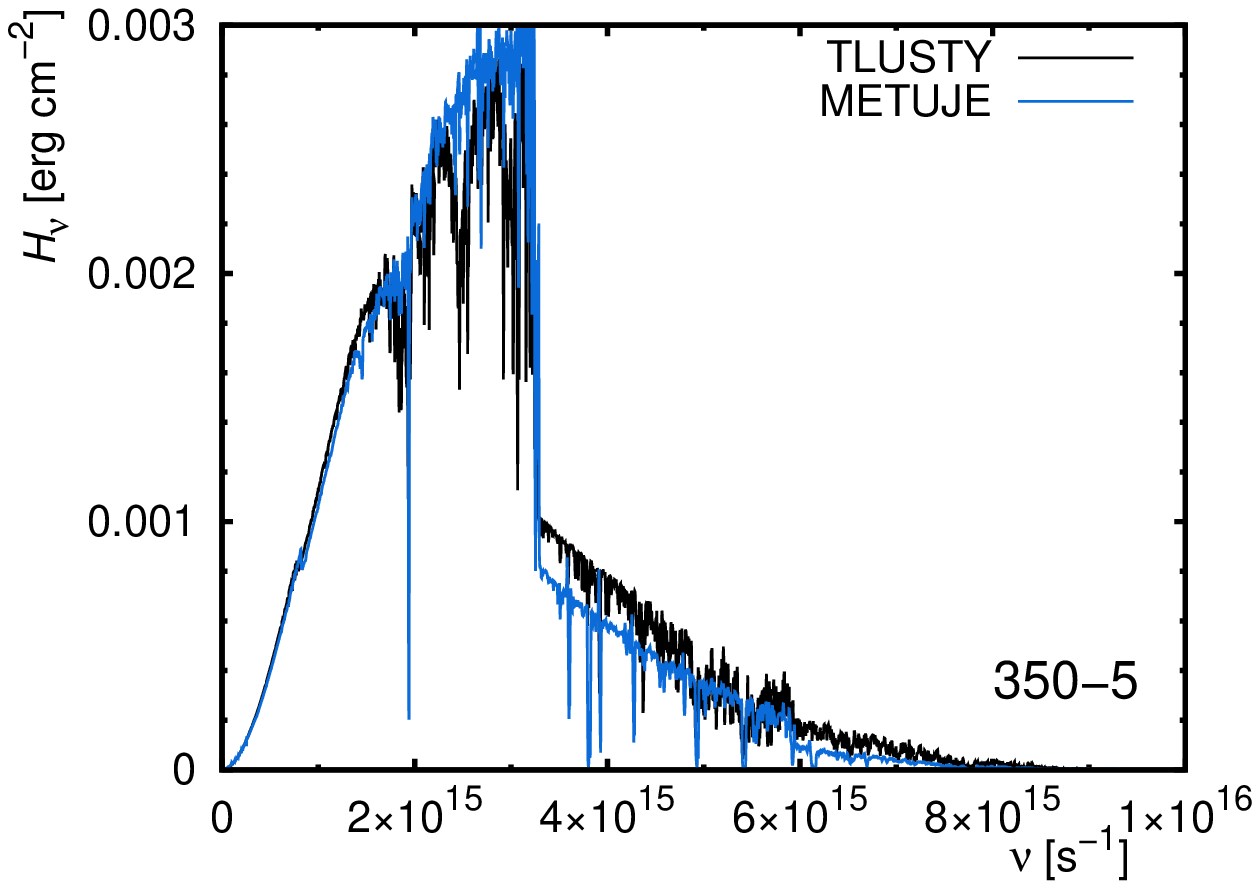}}\\
\resizebox{0.310\hsize}{!}{\includegraphics{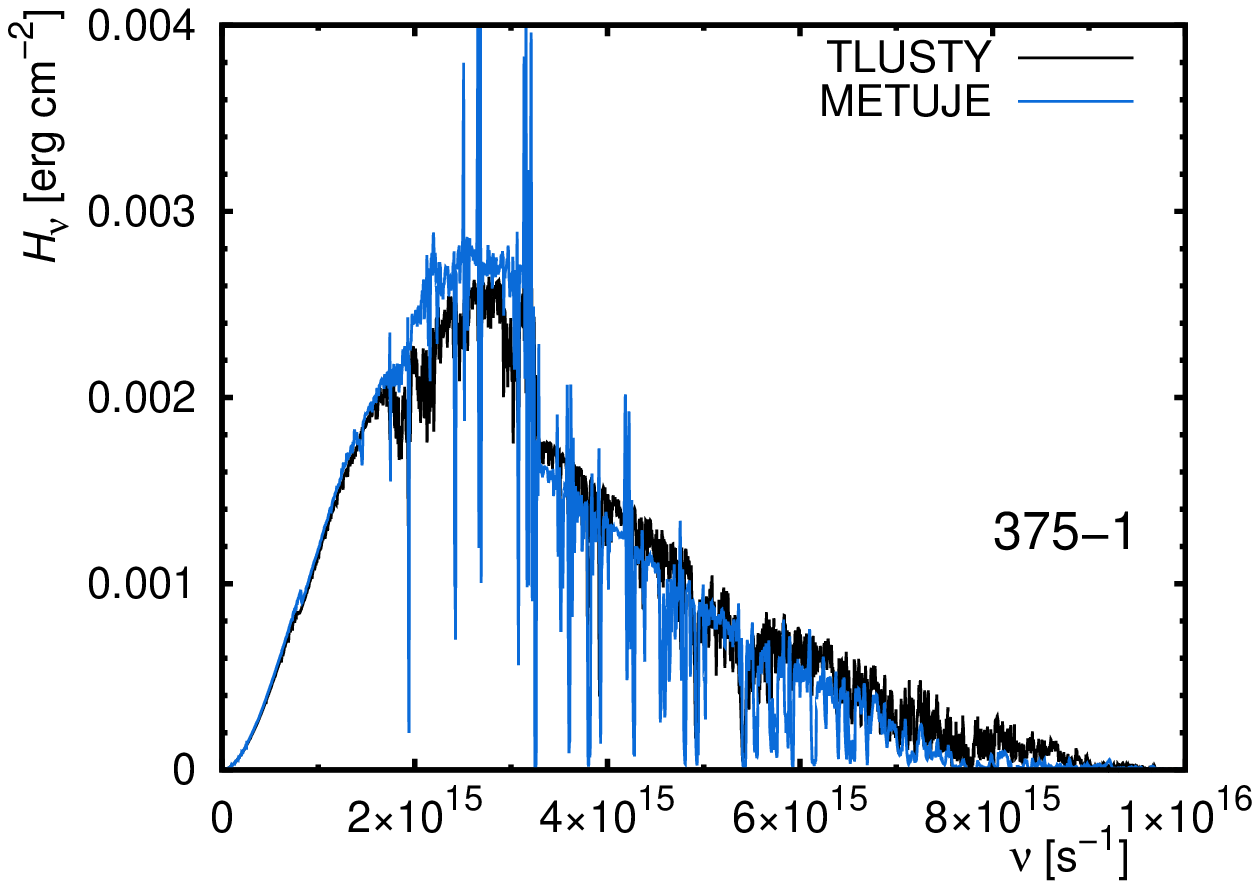}}
\resizebox{0.310\hsize}{!}{\includegraphics{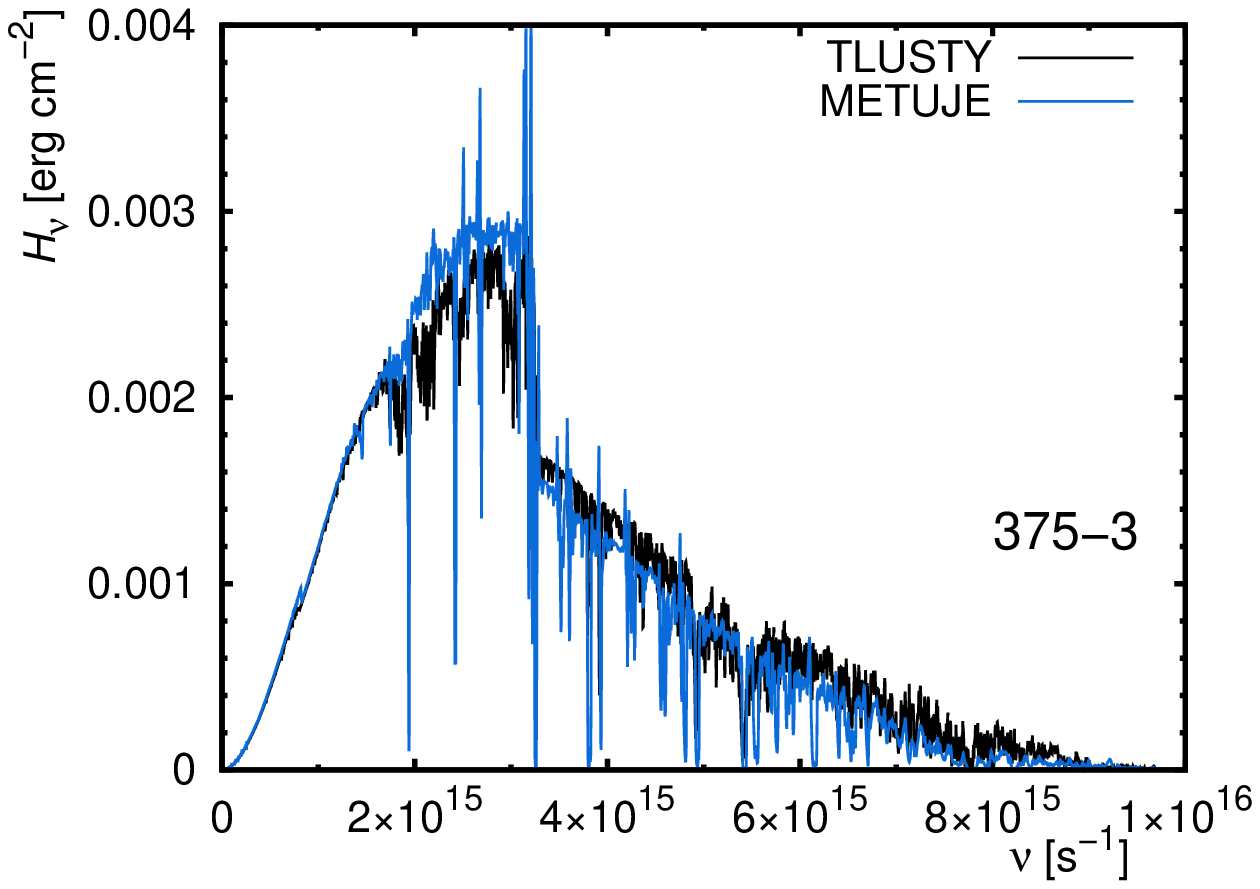}}
\resizebox{0.310\hsize}{!}{\includegraphics{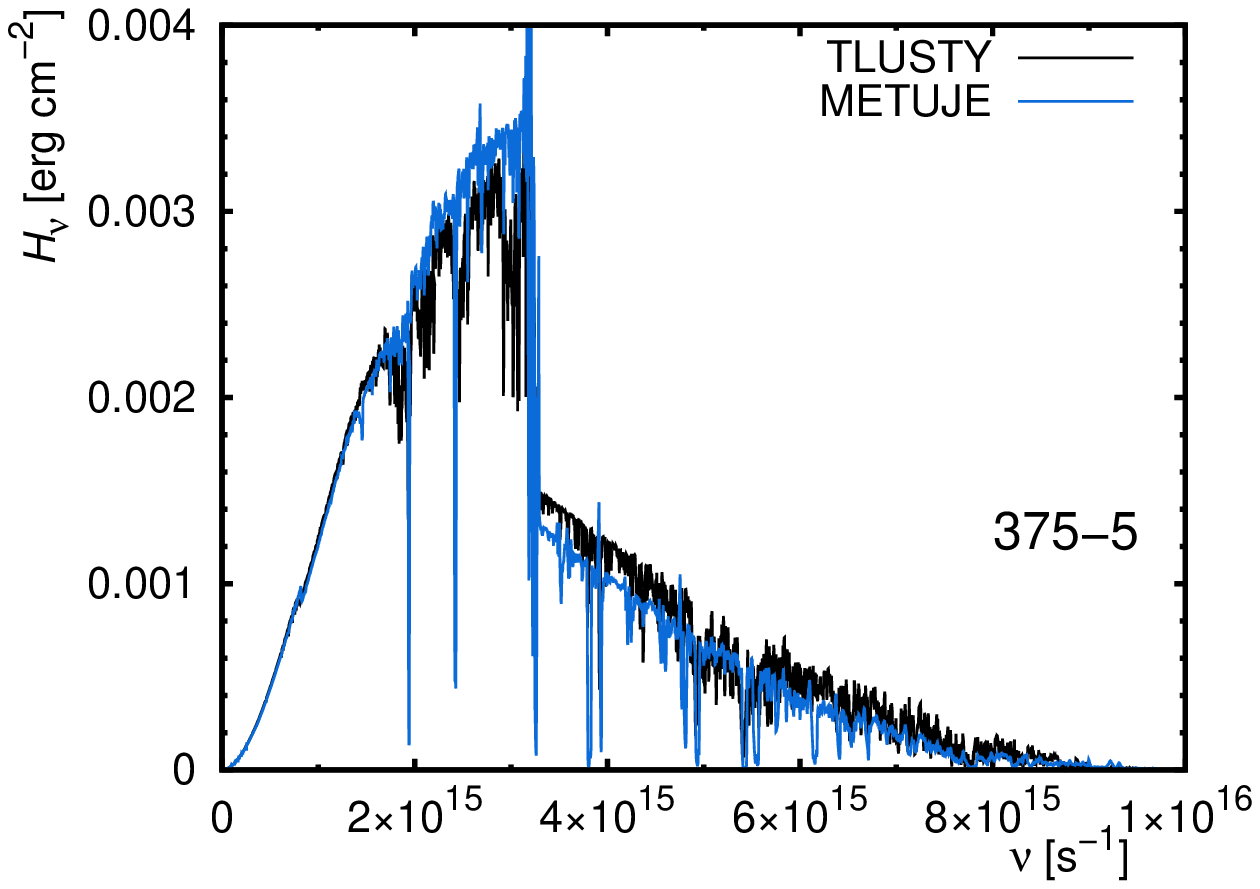}}\\
\resizebox{0.310\hsize}{!}{\includegraphics{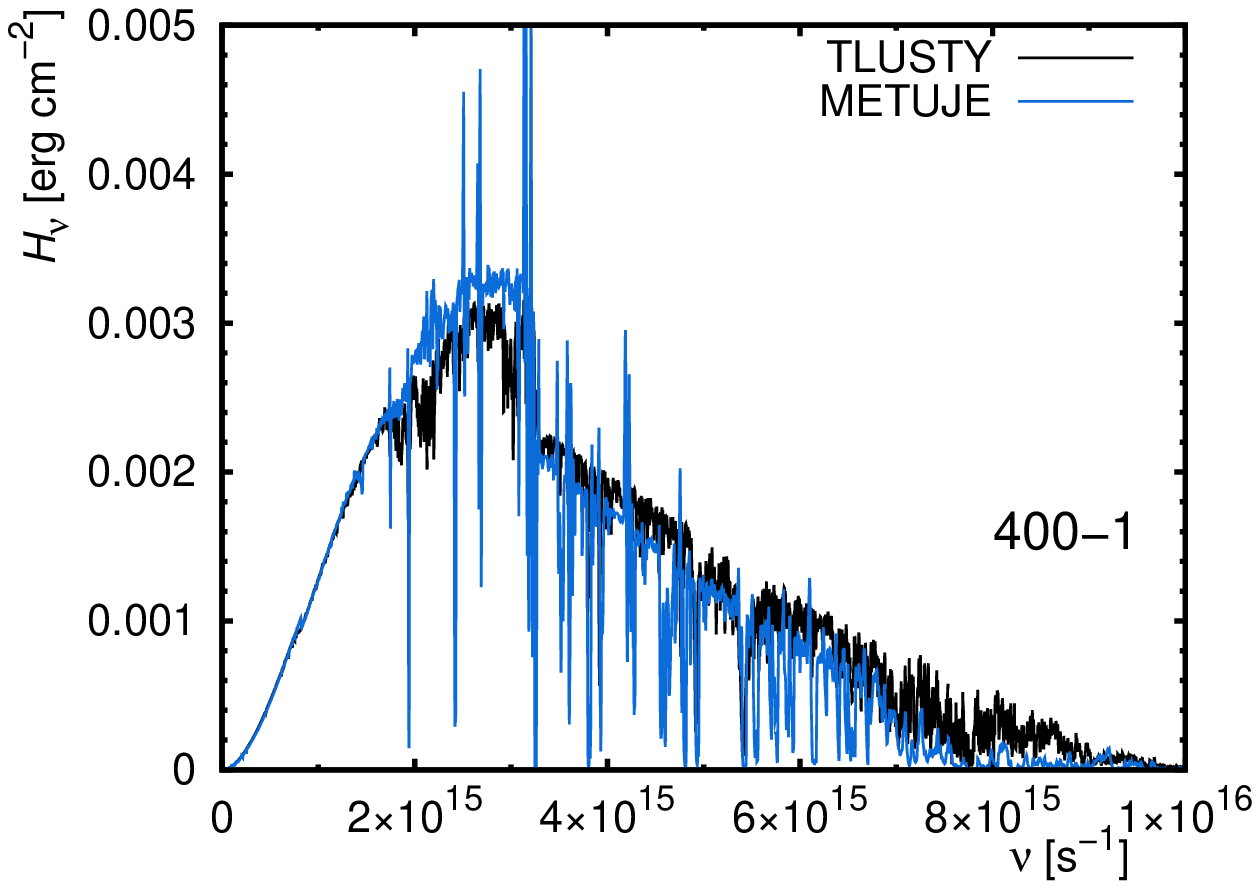}}
\resizebox{0.310\hsize}{!}{\includegraphics{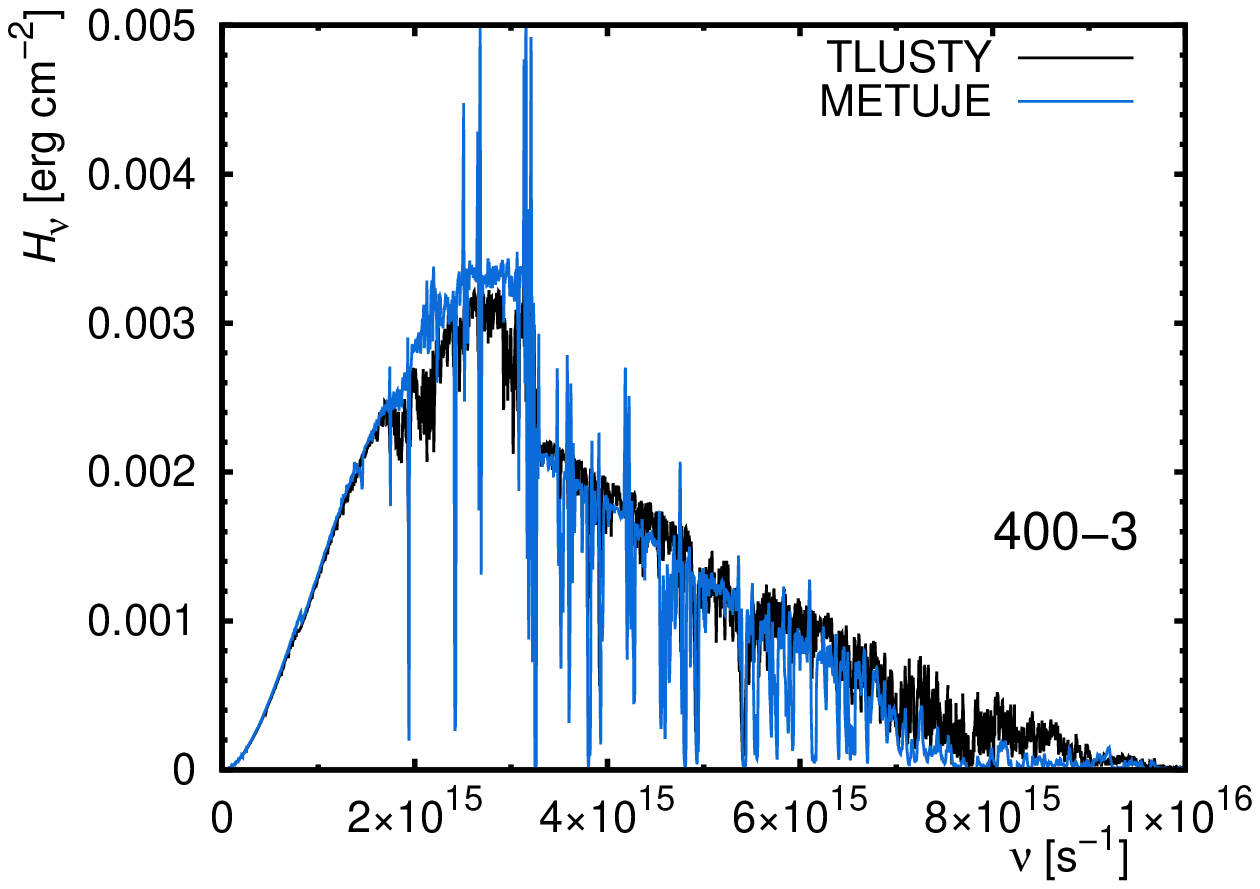}}
\resizebox{0.310\hsize}{!}{\includegraphics{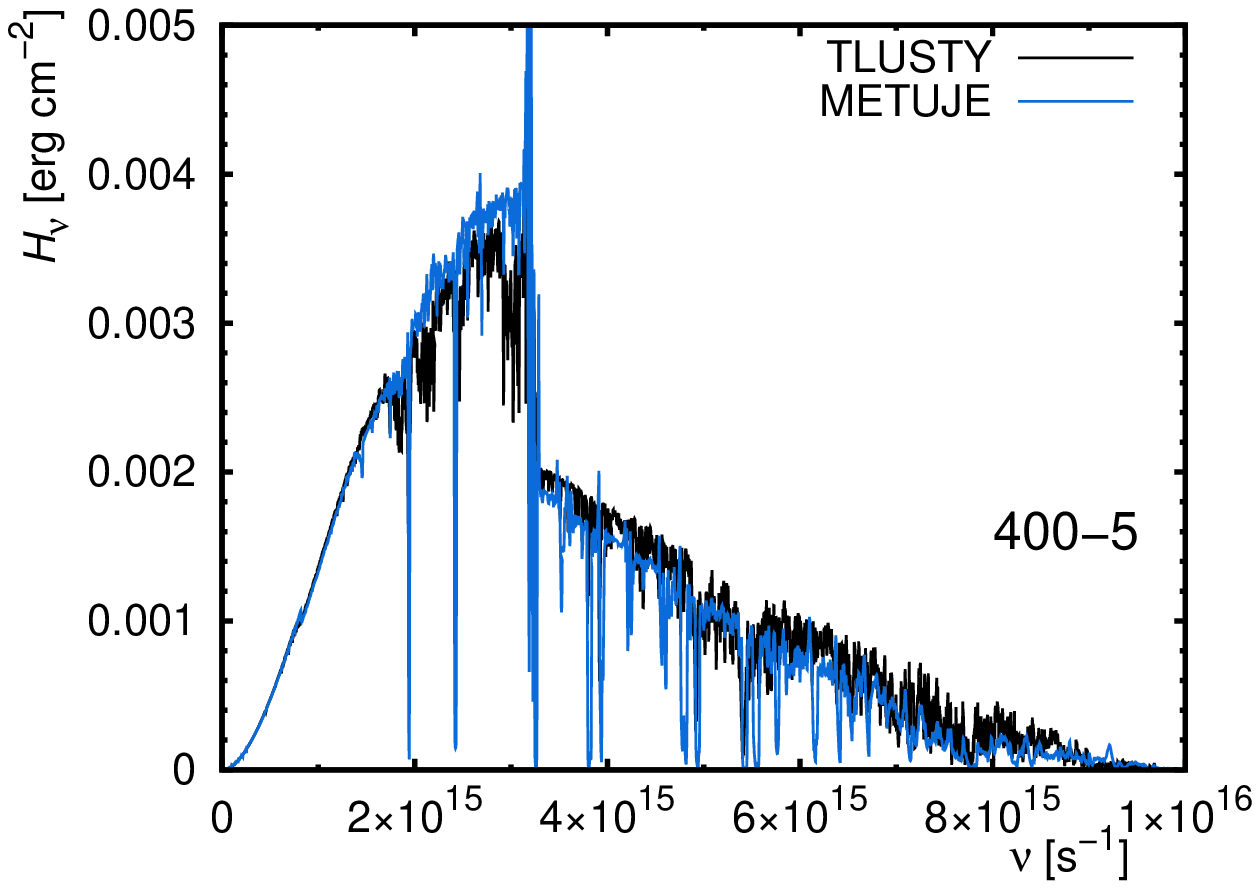}}\\
\resizebox{0.310\hsize}{!}{\includegraphics{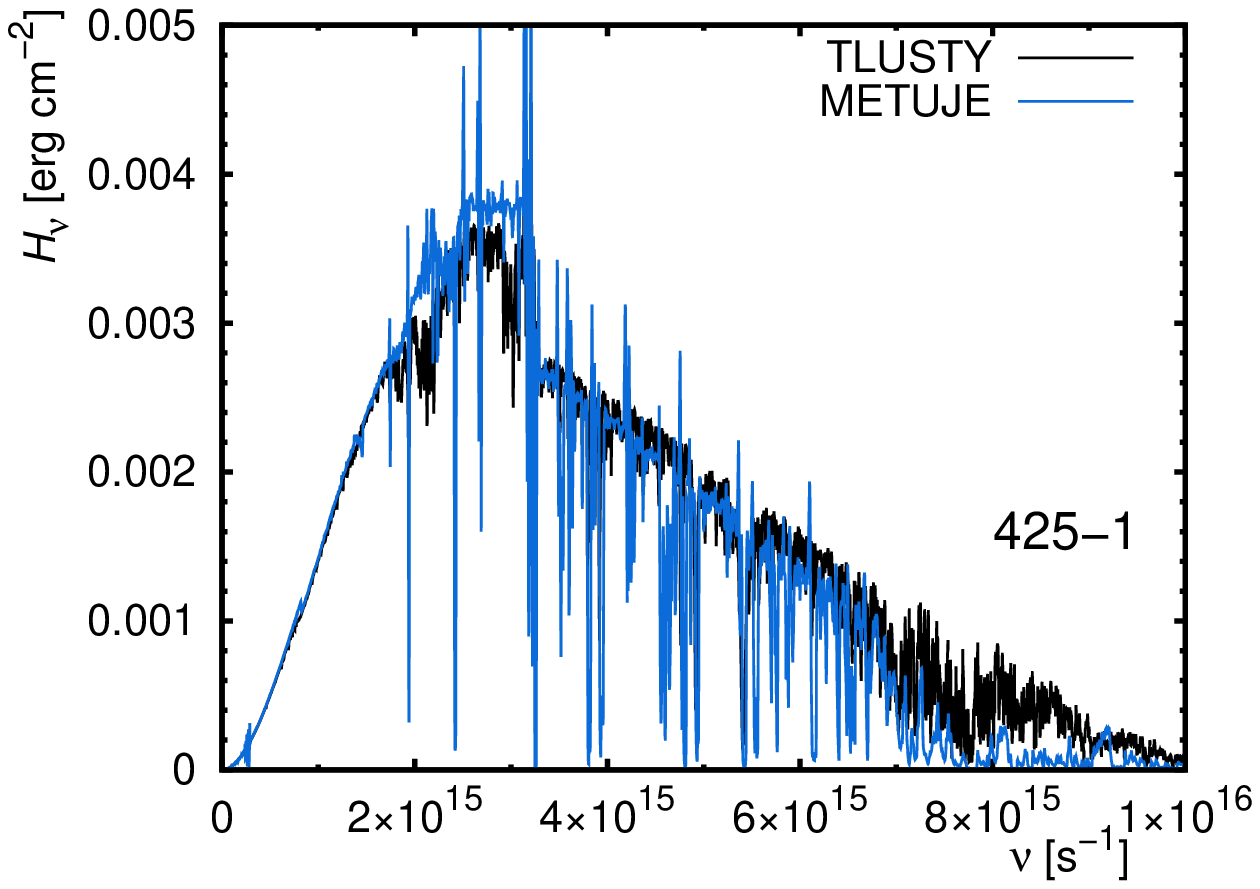}}
\resizebox{0.310\hsize}{!}{\includegraphics{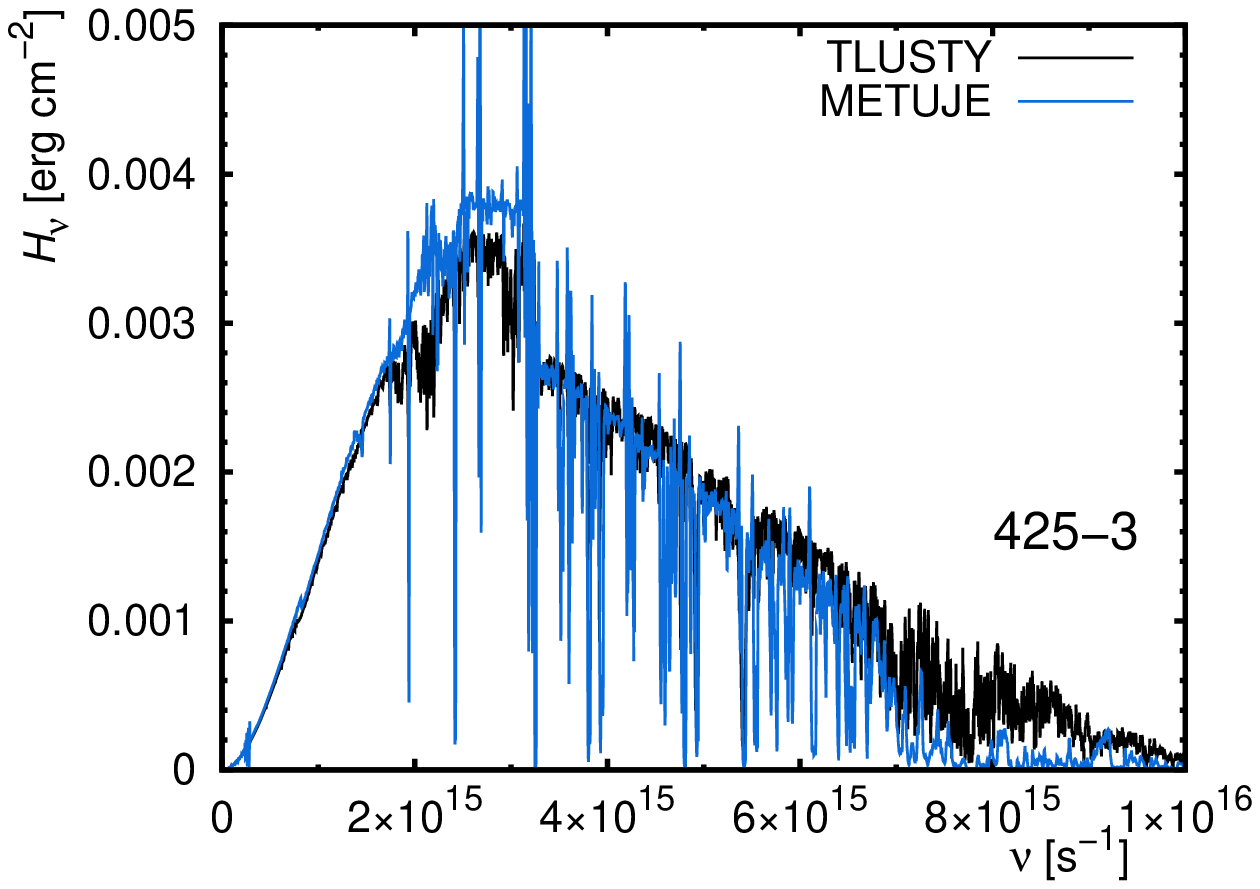}}
\resizebox{0.310\hsize}{!}{\includegraphics{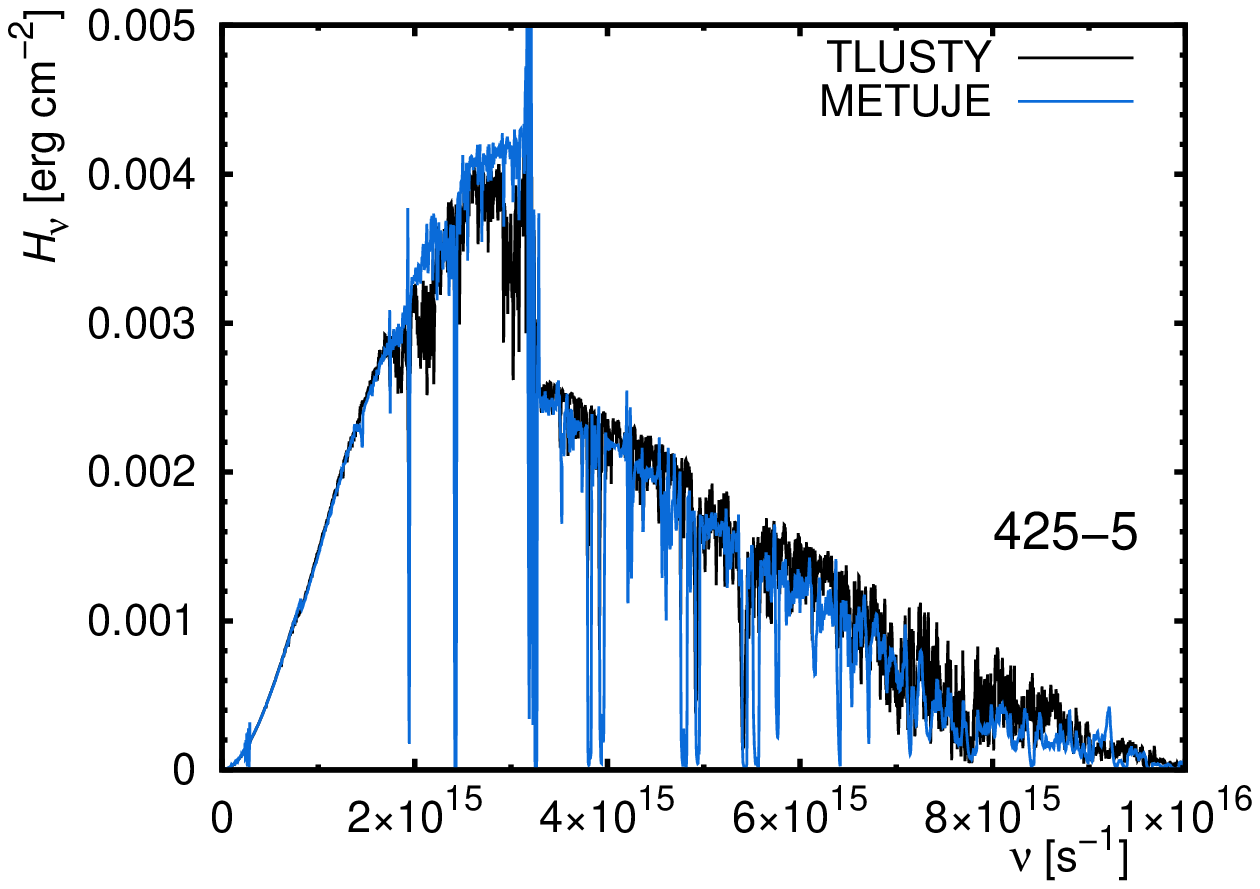}}
\caption{Comparison of the emergent flux from TLUSTY and METUJE models
smoothed by a Gaussian filter. The
graphs are plotted for individual model stars from Table~\ref{ohvezpar} (denoted
in the plots).}
\label{tokmetlu}
\end{figure*}

\section{Calculated wind models}

We calculated a grid of global wind models for stellar parameters corresponding
to O stars in the effective temperature range $30\,000-42\,5 00\,\text{K}$ (see
Table~\ref{ohvezpar}). Stellar masses and radii were calculated using relations
of \citet{okali} for main sequence stars, giants, and supergiants. We assumed
solar chemical composition \citep{asp09} for our models.

\subsection{Comparison with TLUSTY}

TLUSTY\footnote{\url{http://nova.astro.umd.edu/}} \citep{tlusty0, tlustya,
tlusty1} is a computer code for calculation NLTE line-blanketed static
plane-parallel model atmospheres in hydrostatic and radiative equilibrium.
However, it is also often used for analysis of massive stars with winds
\citep[see, e.g.,][and references therein]{boutlucmf, ssh}. In these stars it is
usually applied to their photospheres, where exact treatment of NLTE line
blanketing is an important issue. In this section we compare results of our
calculations with model atmospheres calculated by the TLUSTY code to ensure
basic correctness of our new models.

In Fig.~\ref{tepmetlu} we compare the derived temperature structure of our
models with results of TLUSTY \citep{ostar2003,bstar2006} hydrostatic
plane-parallel model atmospheres calculated for the parameters (effective
temperature, surface gravity, and abundances) corresponding to our dynamic
models. The temperature is plotted as a function of electron number density
$n_\text{e}$, which seems to be a more suitable variable for comparison of
static planparallel and spherical extended dynamic models than the optical depth
or column mass. The temperature structure derived from the METUJE code mostly
nicely agrees with the temperature structure of TLUSTY models in deep
atmospheric layers. Very small differences
that
are apparent especially in cooler models
can be attributed to restricted iron line list in METUJE models. The atmosphere
starts to expand in the low-density regions of Fig.~\ref{tepmetlu} (roughly for
$n_\text{e}=10^{11}-10^{12}\,\text{cm}^{-3}$) resulting in larger differences
between the models. In some models, a typical temperature bump caused by NLTE
effects is apparent \citep{tlusty1, puluni}.

Emergent fluxes from our models also agree reasonably well with results of the TLUSTY
code (see Fig.~\ref{tokmetlu}). The fluxes from the METUJE models were
multiplied by a factor $(R_\text{NR}/R_*)^2$, where $R_\text{NR}$ is the outer
radius of the METUJE models, to obtain the same scale of fluxes. The fluxes nicely
agree especially at lower frequencies. The flux from METUJE models is
significantly lower than the flux from TLUSTY models for frequencies typically
above $7\times10^{15}\,\text{s}^{-1}$ as a result of blocking of the emergent
flux by the wind. The METUJE spectrum is generally flatter; lower flux at high
frequencies is accompanied with higher flux at visual frequencies. This trend
has already been described for simple static spherically symmetric model atmospheres
calculated without line blanketing \citep{kuhumi, kubii}. Small differences at
higher frequencies are also connected with more elaborate line blanketing in the
TLUSTY models, and with simplified treatment of the line source function in the
kinetic equilibrium equations using Sobolev approximation in our models.
Moreover, emission lines originating in the stellar wind appear in the emergent
fluxes. The emergent fluxes from the global models also agree with results of
\citet{pahole} for frequencies lower than the \ion{He}{ii} ionization edge.
Consequently, the METUJE models provide a reasonable radiation field which is
crucial for the wind modeling.

\subsection{Mass-loss rates}

Wind mass-loss rates scale mostly with the stellar luminosity and to a lesser
extent depend on other stellar parameters \citep{cak,vikolamet}. This dependency
is also present in results of our calculations. We found that the predicted
mass-loss rates very tightly correlate with the stellar luminosity. Therefore,
we fitted the mass-loss rates predicted by our models (given in
Table~\ref{ohvezpar}) as
\begin{equation}
\label{dmdtsb}
\log\zav{\frac{\dot M}{1\, \msr }}= - 5.69 + 1.63
\log\zav{\frac{L}{10^6L_\odot}},
\end{equation}
which provides a
relatively accurate approximation for our results with a typical
error of about 10 -- 20 \%.

Predicted mass-loss rates from our global models agree with those derived from
our older purely wind models with fixed atmospheric structure \citep{dusik} for
stars with low luminosity and low mass-loss rates $\dot M\lesssim10^{-7}\,\msr$.
For more luminous stars the global models give mass-loss rates roughly a
factor of two lower than the purely wind models.

This difference is partly due to the effect of modified emergent flux from model
atmospheres, because in previous models we used H-He model atmospheres, whereas
here we include also heavier elements. Our wind models with boundary flux taken
from the model atmosphere with heavier elements predict lower mass-loss rate
than models with H-He atmosphere fluxes as a boundary condition. This is caused
by the difference of emergent fluxes for frequencies higher above
$7-9\times10^{15}\,\text{s}^{-1}$, for which H-He models predict significantly
higher flux than models that include also heavier elements, where the
corresponding flux is blocked \citep[c.f.,][]{lepsipasam}. This was tested with
models in which the boundary flux was taken from model atmosphere with heavier
elements for frequencies lower than $7\times10^{15}\,\text{s}^{-1}$ and from
H-He model for higher frequencies. Such combination of fluxes leads to nearly
the same mass-loss rate as the models with purely H-He fluxes.

Moreover, the same frequency region is additionally blocked by the wind
blanketing effect (mainly by the line transitions). The strongest blocking
occurs close to the sonic point, but the high-frequency radiation is blocked in
the whole wind. This blocking decreases with the distance from the stellar
photosphere. The stronger blocking is connected with the fact that the wind at a
given radius has higher density and, consequently, higher absorption coefficient
than the hydrostatic photosphere. The blocked radiation appears at longer
wavelengths as continuum and emission line radiation \citep{acko,pahole}. The
effect of wind blanketing is stronger for higher densities, and it is of a
lesser importance for winds with low mass-loss rate. The effect of blocking was
tested in the pure wind models for which we neglected acceleration from all line
transitions with frequencies higher than $7\times10^{15}\,\text{s}^{-1}$ and
with boundary flux taken from the model atmosphere. These models give mass-loss rates which are close to those from global models.

\paragraph {Comparison with other predicted mass-loss rates} Our derived
mass-loss rates are significantly lower than those predicted by
\citet{vikolamet}. The difference increases with luminosity and is about a
factor of between two and five for stars with larger luminosity and using \citet{angre} solar
abundances for the calculation of \citet{vikolamet} rates. The difference is
about a factor of between two and four when accounting for lower solar metalicity \citep[see the
rates in Table~\ref{ohvezpar}]{asp09}. Our models show similar behavior when
compared to models of \citet{pahole}.

On the other hand, our predicted mass loss rate for $\zeta$~Pup is by a factor
of about 1.9 higher than the prediction of \cite{gratub}. This difference can be
explained \citep[see][]{cmf1} as a result of strong turbulent broadening
introduced in the models of \cite{gratub}. Moreover, the mass-loss rate for
$\zeta$~Pup predicted by \citet{lepsipasam} using realistic stellar emergent
spectrum and Sobolev approximation is by a factor of $2.4$ higher than our
predicted value for the same parameters. This could be understood as a result of
the multiline effects \citep{pulpren} that cause about a factor of $2.5$
decrease of the predicted mass-loss rate \citep{cmf1}. Our predicted mass-loss
rates also reasonably agree with the predictions of \citet{pulpren}.
\citet{powrdyn} use self-consistent hydrodynamical PoWR models and for
$\zeta$~Pup they derive the mass-loss rate $\dot M=1.6\times10^{-6}\,\msr$,
which agrees reasonably well with our prediction $\dot M=1.2\times10^{-6}\,\msr$
(derived from Eq.~\eqref{dmdtsb} for the same luminosity) especially taking into
account their adopted clumping factor $C_\text{c}=10$.

It is not clear what exactly causes such large differences between our results
and the predictions of \citet{vikolamet} and \citet{pahole}. Although the physics involved is the same, there are differences in the solution of the radiative
transfer equation, in treatment of the hydrodynamics, and in derivation of the
wind mass-loss rate. The most significant difference is connected to the
calculation of the radiative force, especially close to the star. The models
that predict large mass-loss rates in comparison to ours use the Sobolev
approximation to calculate the radiative force \citep{vikolabis,cinskasmrt}
throughout the wind. Also our models with the Sobolev approximation
\citep{nltei} predict significantly higher mass-loss rates than the consistent
CMF models presented here. Moreover, our models and the models of \cite{gratub}
use the radiative force directly in the hydrodynamical equations, whereas
\citet{vikolamet} derive the mass-loss rate from global energy balance and
\citet{pahole} use line force multipliers. On the other hand, while the models
of \citet{vikolamet} use the line list that includes all elements from H to Zn,
we include only elements for which we solve the kinetic equilibrium equations
\citep[see][]{nlteiii}.

\paragraph {Comparison with CMFGEN} The code CMFGEN \citep{hilmi} calculates
consistent NLTE spectra of stellar atmospheres with winds assuming the velocity
structure, which is usually given by a modified $\beta$-velocity law
\citep[e.g.,][]{vesnice}. The code does not solve the hydrodynamic equations. As
a consequence, direct comparison with our results is not possible. However, the
code can be used to calculate the radiative force and to test the consistency of
derived solution. \citet{bouhil} for their model of $\zeta$~Pup (see their
Sect.~6.5) derive the hydrodynamically consistent solution for $\dot
M=2.7\times10^{-6}\,\msr$, for which we predict $\dot M= 1.7
\times10^{-6}\,\msr$. \citeauthor{bouhil} assumed the volume filling factor
$f_\infty= 0.1 $ in their analysis (note that $f_\infty=1/C_c$, the clumping
factor). For a wind model with volume filling factor $f_\infty= 0.1 $ we
predicted an increase in mass-loss rates by a factor of two \citep{sanya} with respect to
a wind model without clumping. A similar situation appears for the model of
\citet[Fig.~18]{vesnice}. Although their model of $\epsilon$~Ori with $\dot
M=4.9\times10^{-7}\,\msr$ shows a slightly insufficient radiative force to drive
the wind close to the star, our prediction gives $\dot M= 4.4
\times10^{-7}\,\msr$ not far from their value. \citet{peta} derive the wind
mass-loss rates of BA supergiants from the requirement of energy conservation
using the CMFGEN code taking into account clumping (actually microclumping). The two
hottest stars from their sample with $T_\text{eff}=30\,000\,$K and different
masses overlap with our sample. For these stars \citeauthor{peta} derive
mass-loss rates $6.7\times10^{-7}\,\msr$ and $3.9\times10^{-7}\,\msr$. This agrees with our predictions $ 8.0 \times10^{-7}\,\msr$ and $ 3.1
\times10^{-7}\,\msr$. From this we conclude that our resulting mass-loss rates
are roughly consistent with radiative force predictions derived from CMFGEN
code.

\paragraph{Comparison with observations}

\begin{figure}[t]
\centering
\resizebox{\hsize}{!}{\includegraphics{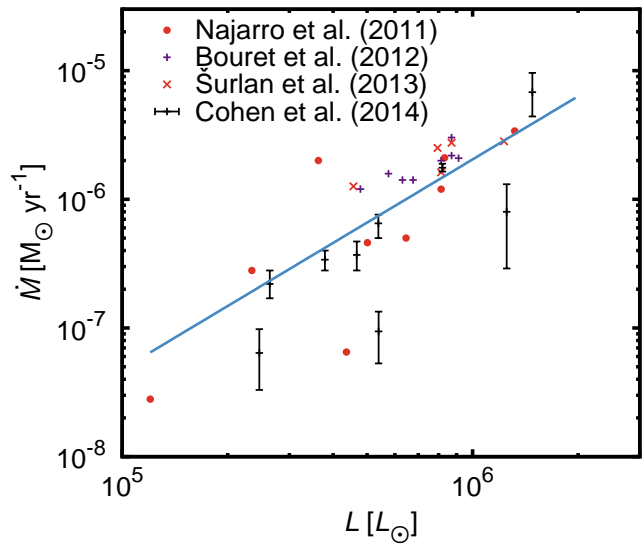}}
\caption{Predicted dependence of the mass-loss rate on luminosity
Eq.~\eqref{dmdtsb} (solid line) in comparison with data derived from infrared
line profiles \citep{najaro}, combined optical and UV analysis
\citep{bouhil,clres2}, and X-ray line profiles \citep{cohcar}.}
\label{dmdtl}
\end{figure}

In Fig.~\ref{dmdtl} we compare the predicted dependence of the mass-loss rate on
luminosity with data derived from near-infrared line spectroscopy
\citep{najaro}, combined optical and UV analysis that accounts for clumping
\citep{bouhil,clres2}, and X-ray line profiles \citep{cohcar}. From the
comparison we excluded the mass-loss rates derived from X-ray line profiles of
HD~93250, which is hotter than stars considered here, and those of $\iota$~Ori and
$\zeta$~Oph, which have extremely low mass-loss rates for their luminosities and
may display the "weak wind problem". When comparing with observational data, we
refrained from using more common modified wind momentum \citep{kupul}. The wind
momentum depends on the wind terminal velocity, which is sensitive to the
ionization equilibrium in the outer parts of the wind. Consequently, the
terminal velocity may be affected by radial variations of clumping or by X-rays
\citep[e.g.,][]{sanya}.

Our predictions agree with the mass-loss rates derived from X-ray line
profiles \citep{cohcar,rahen} that were obtained with low uncertainty and with
mass-loss rates derived from near-infrared emission lines \citep[in combination
with other diagnostics,][]{najaro}. Our predictions are roughly a factor of
1.6 lower than those derived from combined UV and optical analysis
\citep{bouhil,clres2}. The UV line profiles \citep{clres2} indicate that
clumping starts relatively close to the star. If the clumping has already started in
the region where the mass-loss rate is determined, then it may lead to an
increase of the wind mass-loss rate due to enhanced recombination \citep{muij}.
Consequently, the true mass-loss rates may even be slightly higher.

\begin{figure}[t]
\centering
\resizebox{\hsize}{!}{\includegraphics{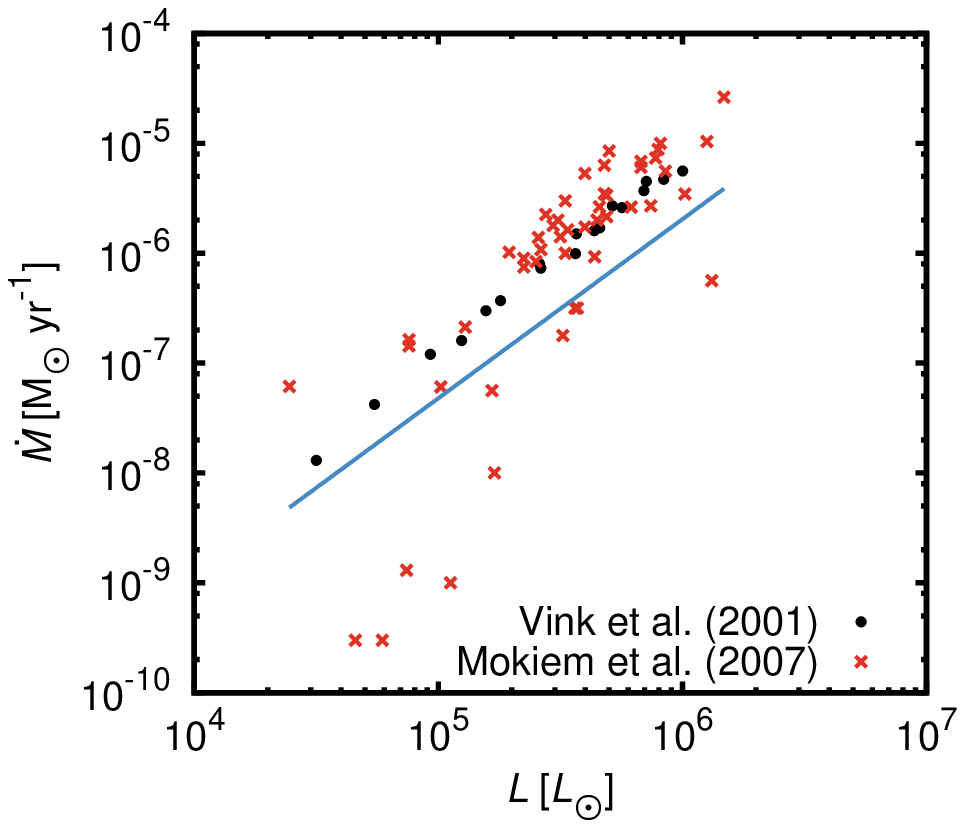}}
\caption{Predicted dependence of the mass-loss rate on luminosity
Eq.~\eqref{dmdtsb} (solid line) in comparison with predictions of
\citet{vikolamet} for stars from Table~\ref{ohvezpar} and H$\alpha$ analysis
compiled by \citet[taken from \citealt{moksez4,moksez2,moksez3}, and
\citealt{moksez1}]{mokz}.}
\label{dmdtv}
\end{figure}

Our predicted mass-loss rates are a factor of $ 4.7 $ lower than unclumped
mass-loss rates derived from H$\alpha$ line (as compiled by \citealt{mokz}, see
Fig.~\ref{dmdtv}). Taking into account clumping, the H$\alpha$ mass-loss rates
would be lower by a factor $C_\text{c}^{1/2}$, where $C_\text{c} = 1/f_\infty$
is the clumping factor. Attributing the difference between our predictions and
compilation of \citet{mokz} to clumping would therefore imply the clumping
factor of $C_\text{c}= 4.7^2=22$. On the other hand, in the case in which the
clumping starts close to the star, the clumping also affects the theoretical
predictions \citep{muij}. For optically thin inhomogeneities (microclumping)
present throughout the whole wind, the mass-loss rate scales on average as $\dot
M\sim C_\text{c}^\alpha$ with $\alpha\approx1/4$ \citep[Table~3]{muij}.
Consequently, if the difference between the predicted and H$\alpha$ mass-loss
rates stems from the influence of clumping on both observations and predictions,
then the required clumping factor is from $C_\text{c}^{1/2}C_\text{c}^{1/4}= 4.7
$ equal $C_\text{c}= 8 $. This would also imply that the true mass-loss rates
are by a factor of $C_\text{c}^{1/4}= 1.7 $ higher than our predicted values and
agree with results of \citet{bouhil} and \citet{clres2}. However, if the
clumping starts above the critical point, the difference between observation and
theory would imply clumping factor $C_\text{c}= 22 $, since the mass-loss rate
is determined in the region below the critical point.

The difference between microclumping and macroclumping is crucial for further
mass-loss rate analysis. For example, the clumping factors adopted by \citet{bouhil}
and which account only for microclumping are somehow larger than those found by
\citet{clres2}, who took a more general model of macroclumping into account. The
solution degeneracy in a clumped wind with interclump media may complicate the
empirical mass-loss rate determination from UV lines \citep{clres1,supujo}.
Moreover, a detailed multiwavelength analysis of $\delta$~Ori~Aa1
\citep{shendelori} gives a mass-loss rate which is about a factor of three higher than
the value predicted here and which agrees with predictions of \citet{vikolamet}.

\subsection{Terminal velocities and ionization fractions}

\begin{figure*}[t]
\centering
\resizebox{0.32\hsize}{!}{\includegraphics{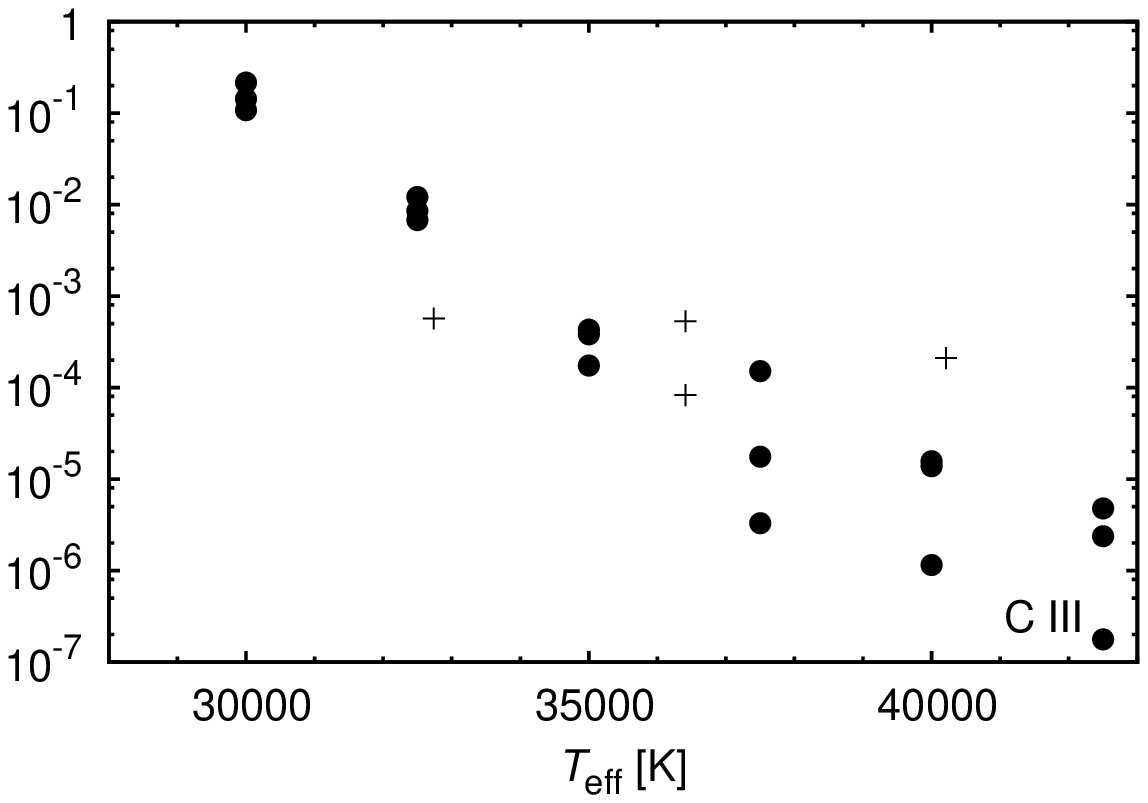}}
\resizebox{0.32\hsize}{!}{\includegraphics{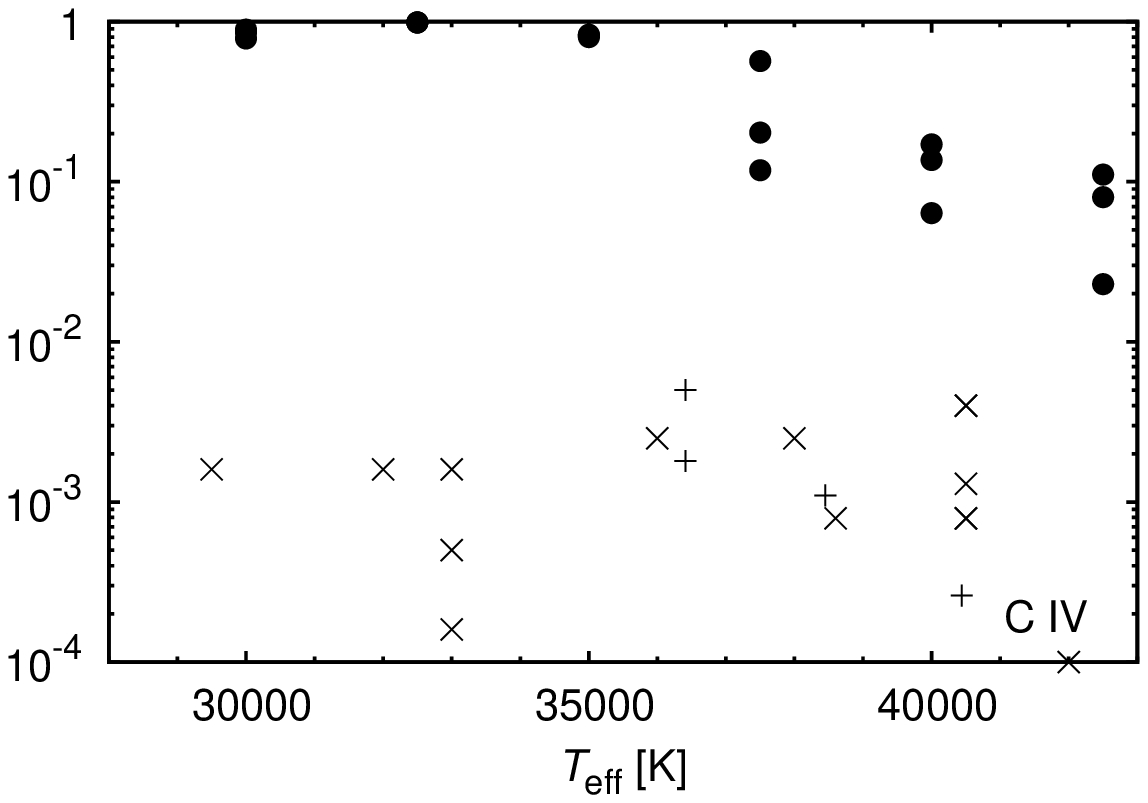}}
\resizebox{0.32\hsize}{!}{\includegraphics{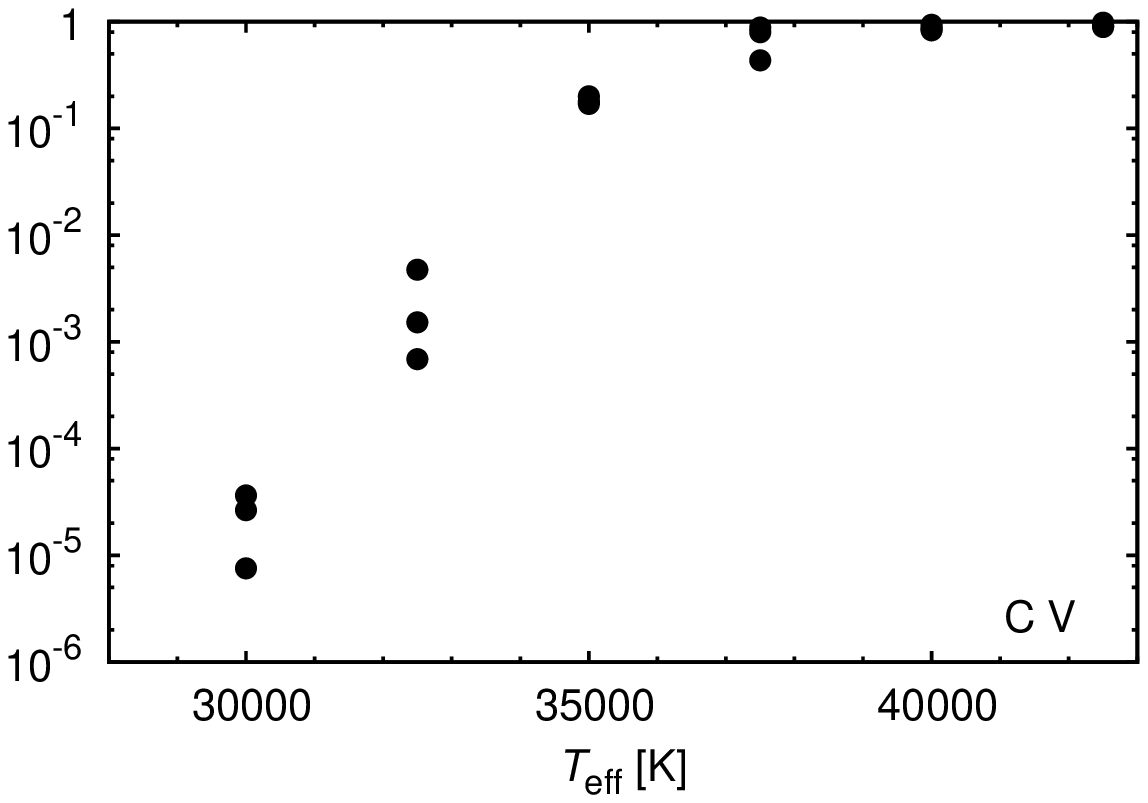}}
\resizebox{0.32\hsize}{!}{\includegraphics{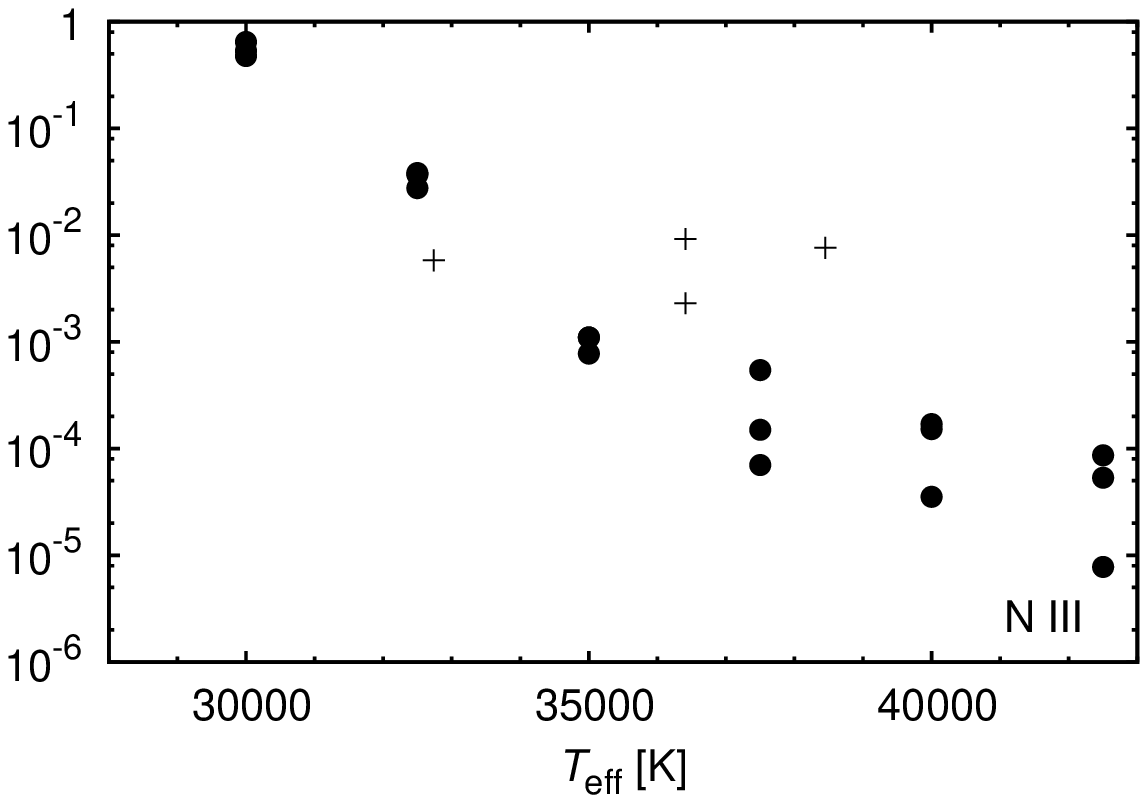}}
\resizebox{0.32\hsize}{!}{\includegraphics{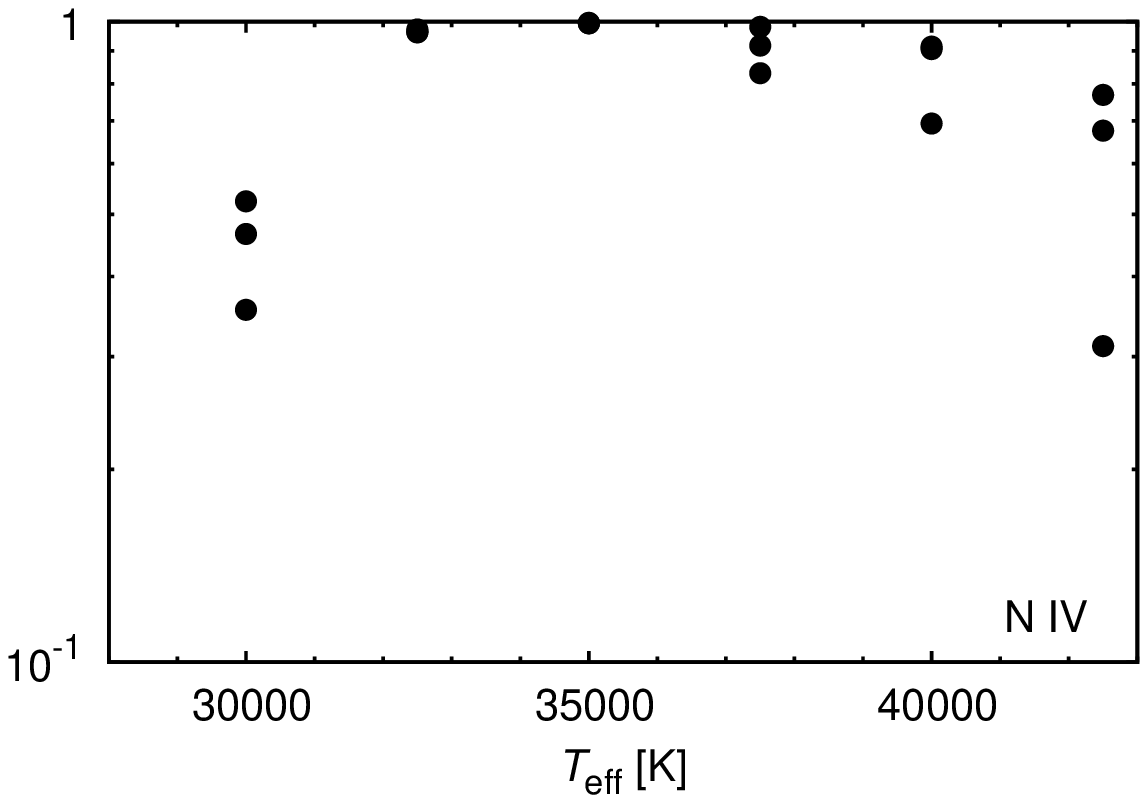}}
\resizebox{0.32\hsize}{!}{\includegraphics{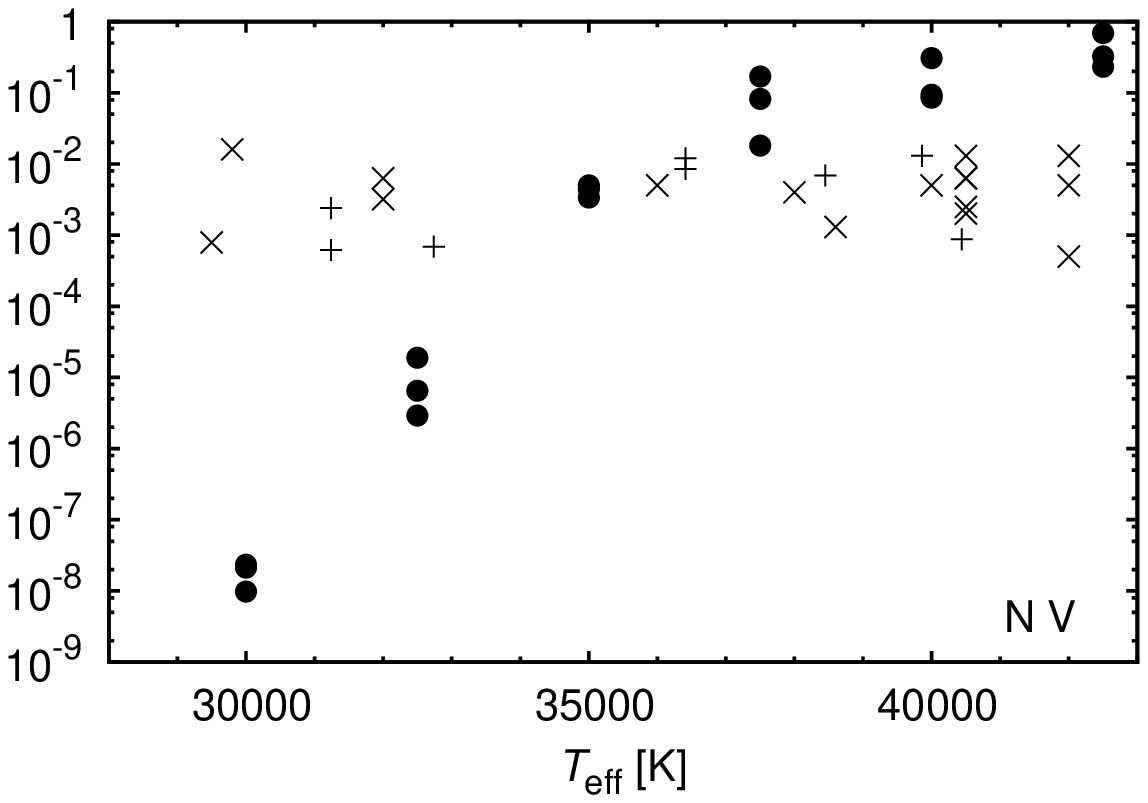}}
\resizebox{0.32\hsize}{!}{\includegraphics{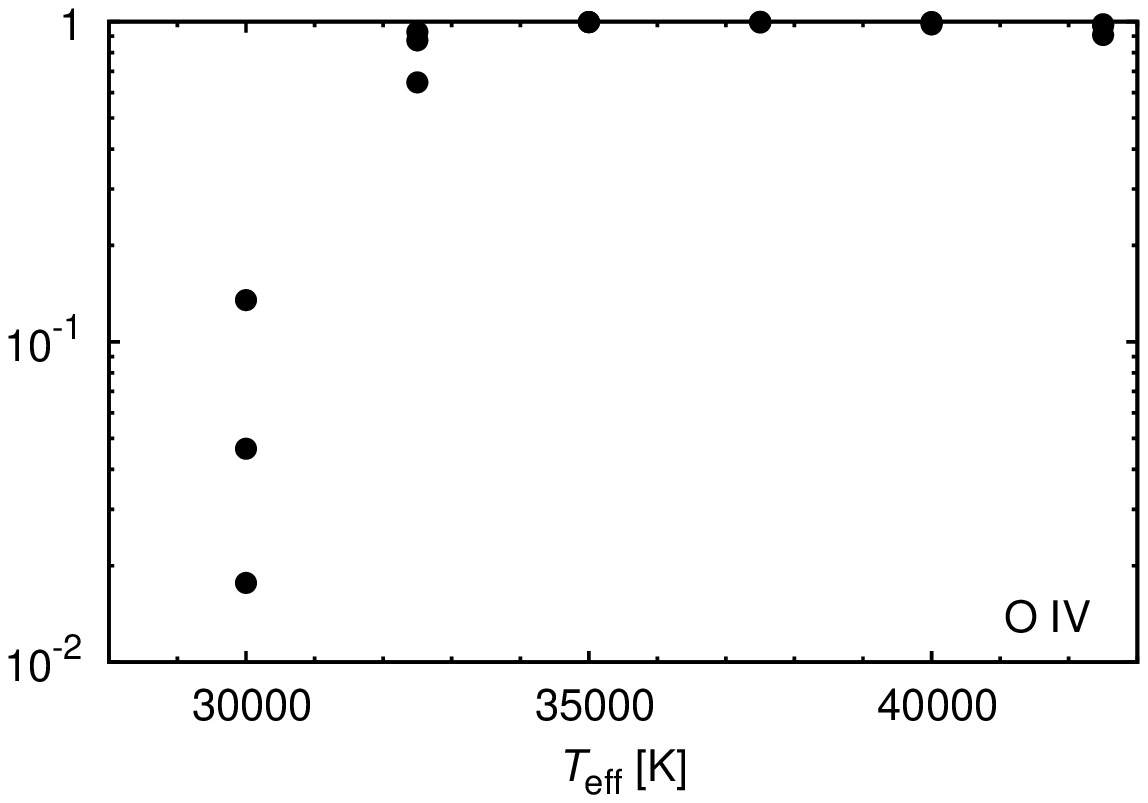}}
\resizebox{0.32\hsize}{!}{\includegraphics{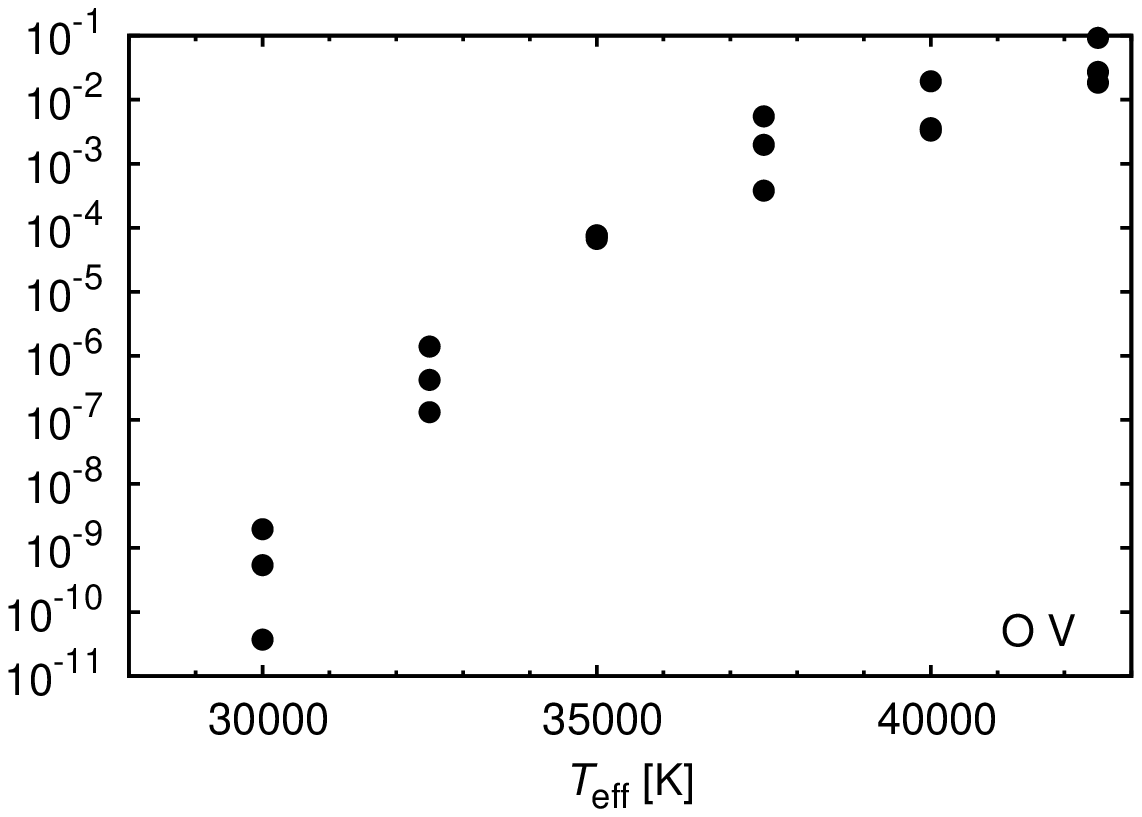}}
\resizebox{0.32\hsize}{!}{\includegraphics{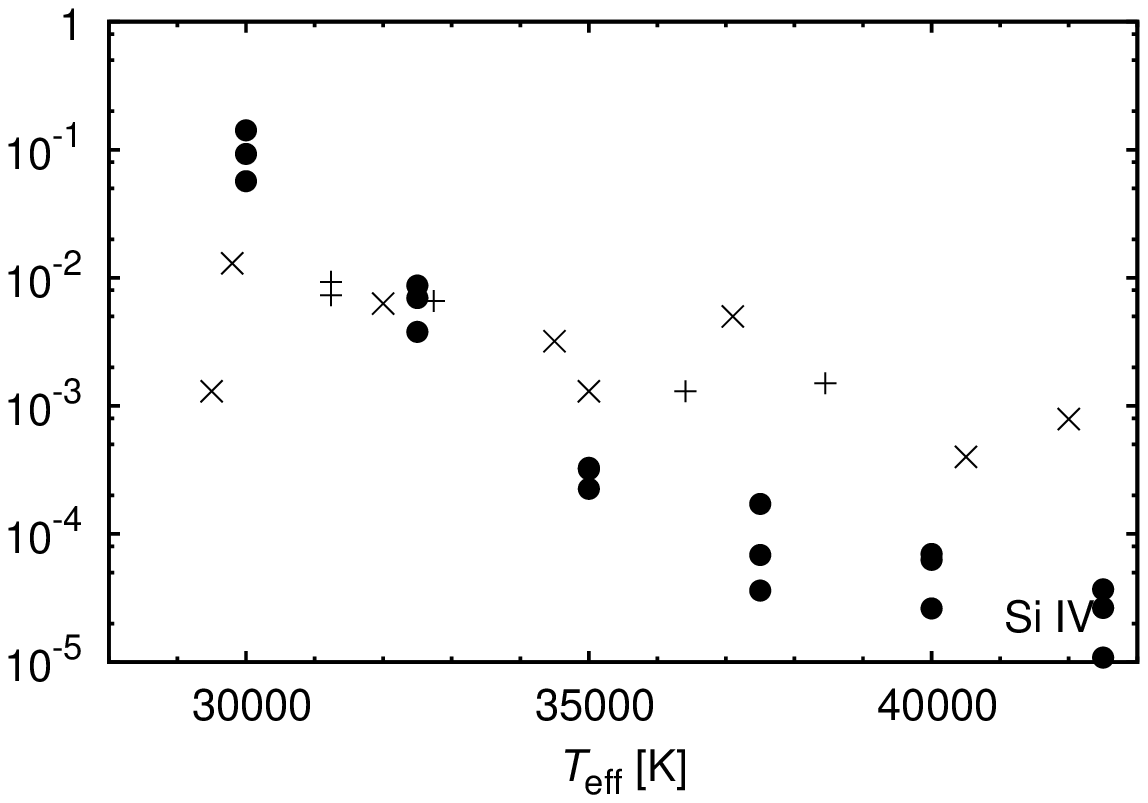}}
\resizebox{0.32\hsize}{!}{\includegraphics{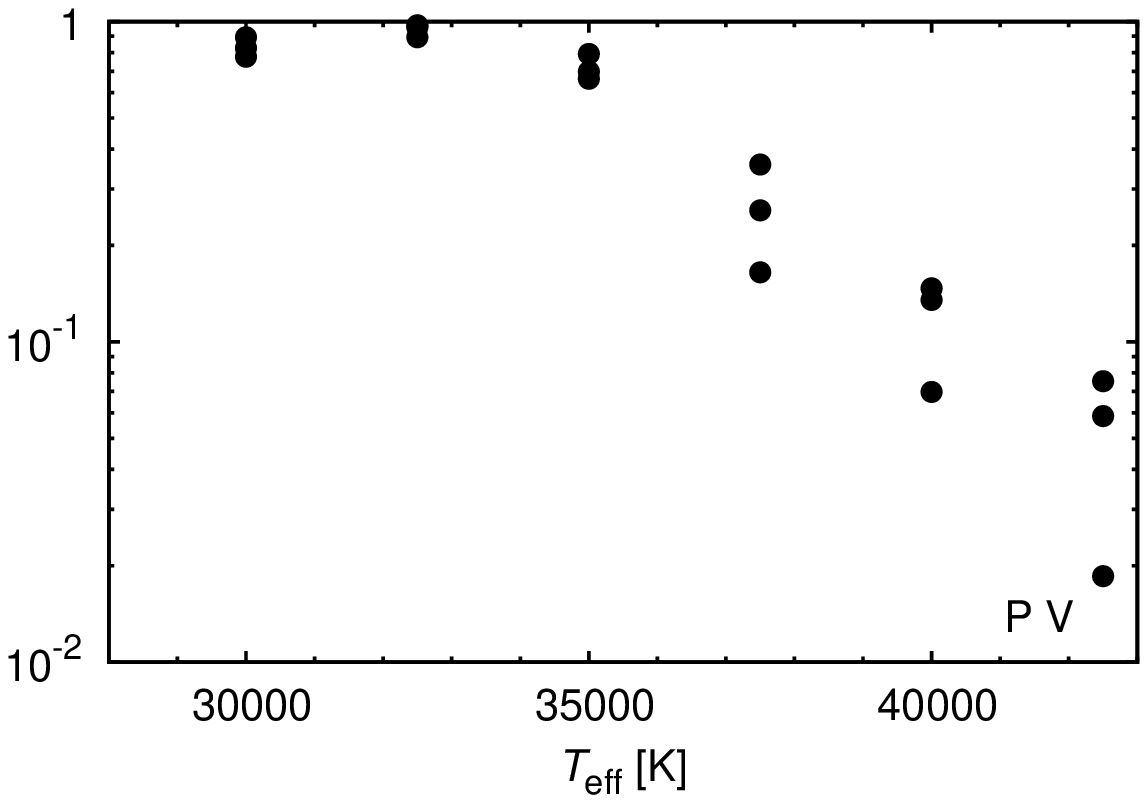}}
\resizebox{0.32\hsize}{!}{\includegraphics{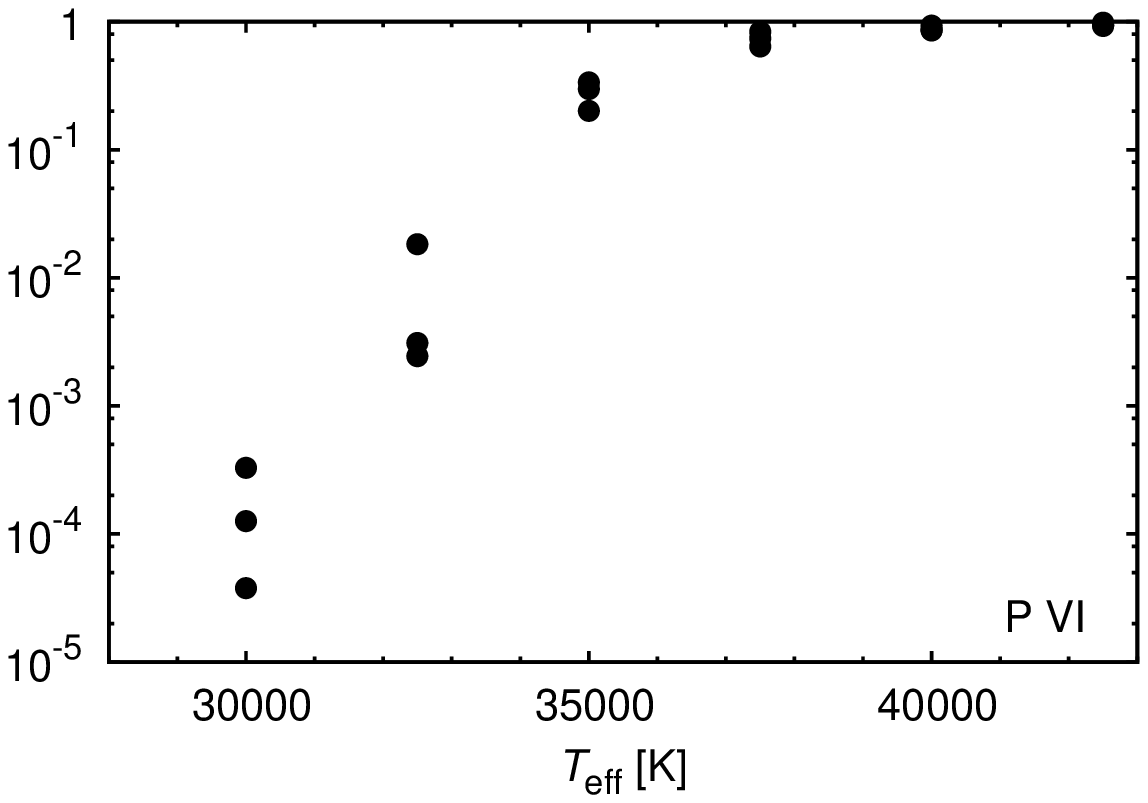}}
\caption{Ionization fractions as a function of the effective temperature for
individual stars from our sample at the point where the radial velocity reaches
half of its terminal value. The ionization fraction derived from observations
were adopted from \protect\citet[for LMC stars, plus signs $+$]{maso} and
\protect\citet[only for a representative sample of stars, crosses
$\times$]{hp}.}
\label{iontep}
\end{figure*}

From the theoretical considerations it follows that the wind terminal velocity
$v_\infty$ is proportional to the escape speed $v_\text{esc}$ \citep{cak}.
\citet{lsl} derived an average ratio $v_\infty/v_\text{esc}=2.6\pm0.2$ for
Galactic O stars. Our models predict slightly lower ratio
$v_\infty/v_\text{esc}= 1.9 $. The difference could possibly be caused by
clumping, which we did not consider in our models. Observational analysis
typically assumes that clumping increases with radius close to the star
\citep[e.g.,][]{najaro,bouhil}. Enhanced clumping in the region above the
critical point causes increase of the terminal velocity \citep{sanya}.

In Fig.~\ref{iontep} we plot the ionization fraction of selected ions at radius
where $v \approx \vinfty/2$ as a function of the stellar effective temperature.
In our calculations we do not consider any source of enhanced X-ray radiation,
which may cause X-ray ionization \citep{casol,macown,pahole,nlteiii,lojza} and
we also do not consider clumping \citep{potclump,muij}. These effects may
influence the comparison of our results with observations. Nonetheless, except for
that of \ion{C}{vi,} the ionization fractions agree fairly well with observations. Moreover,
our calculated ionization fractions agree with those ionization fractions
calculated by \citet[Fig.~8]{lojza} which ignore the X-ray ionization. 

\section{Discussion and conclusions}

We provide global (unified) wind models of O stars, which account for the
stellar photosphere and wind. The models are obtained by simultaneous solution
of hydrodynamic, radiative transfer, and kinetic equilibrium (NLTE) equations
throughout the photosphere and wind. We provide a formula that predicts the wind
mass-loss rate as a function of stellar luminosity for O stars with solar
chemical composition.

As found by \cite{bouhil} and \cite{vesnice}, some calculations with
prespecified wind velocity law predict too low radiative acceleration around the
sonic point. However, the prespecified velocity law does not necessarily fulfill
the momentum equation. On the other hand, our models smoothly pass from the
stellar photosphere in hydrostatic equilibrium to the fast radiativelly
accelerated wind. This shows that consistent wind models that properly treat the
momentum equation at the wind base are able to solve the problem of
hydrodynamical consistency.

We compare our models with other models available in literature and with
observations. Our models provide very similar atmospheric structure to TLUSTY
models deep in the stellar atmosphere. The emergent flux from our models also
reasonably agrees with TLUSTY models, and the differences between both models
can be attributed mostly to the wind blanketing and to the effects caused by
differences between plane-parallel and spherically symmetric model atmospheres
\citep{kuhumi, kubii}.

With respect to our previous models that did not include the photosphere
consistently, the derived mass-loss rates are on average lower by a factor of two.
This is caused by stronger blocking of the radiative flux for frequencies higher
than about $7\times10^{15}\,\text{s}^{-1}$, for which the stellar wind is still
optically thick for lower velocities.

Our models provide significantly lower mass-loss rates than the models of
\citet{pahole} and \citet{vikolamet}. Both these latter models use the Sobolev
approximation for the calculation of the radiative force, which is not
appropriate close to the star and in the case of line overlaps. Moreover,
neither of these models solves the radiation hydrodynamics completely
selfconsistently. \citet{vikolamet} use a prespecified hydrodynamical structure
and a global energy balance to derive the mass-loss rate and \citet{pahole} use
line force multipliers to calculate the radiative force. Calculation of
\citet{divnetomaj} which aims to provide hydrodynamic models with Monte Carlo
radiative transfer, but with the Sobolev approximation, still provides mass-loss
rate of an OV star which is about three times higher than our prediction. The
Sobolev approximation provides reliable results for the calculation of radiative
force coming from single line in supersonic wind \citep{cmf1}, but the simple
single line Sobolev approximation that neglects continuum fails to treat
multiline effects, influence of the lines on continuum, and does not provide
reliable results in the vicinity of the sonic point. We conclude that the most
probable reason for the lower mass-loss rate obtained by our models is an improved
calculation of the line force, which is done using the CMF approach, in contrast
to the Sobolev approximation, which was used in studies mentioned here.

On the other hand, our predicted mass-loss rate for $\zeta$~Pup
is slightly higher than that predicted by hydrodynamic calculations of
\citet{gratub} using the CMF line force. Moreover, our calculations are
consistent with the radiative force derived from the CMFGEN code, which also
uses the more general CMF method of solution of the radiative transfer equation.

The radiative force is proportional to the product of the flux and opacities.
Consequently, all differences in the mass-loss rate should be attributed either
to the differences in fluxes or opacities at last. Our detailed comparison with
fluxes presented in \citet{pahole} has not revealed any significant differences
between the models that could cause a significant difference between predicted
mass-loss rates. The difference in opacities may be caused either by inaccurate
level populations or by missing opacities. However, we found reasonable
agreement with ionization fractions calculated by \citet{lojza}. Our tests using
line data from different sources have not revealed any significant opacity gap
for lighter elements with $Z\leq20$. There are some missing iron lines in the
list, however their inclusion has not led to a significant increase of the
radiative force. Consequently, there is likely a different reason for the
difference of mass-loss rate between individual models.

Our derived mass-loss rates agree with observational results corrected for
clumping. The predicted mass-loss rates agree nicely with mass-loss rates
derived from X-ray line profiles \citep{cohcar} and based on near-infrared lines
\citep{najaro}. Our derived mass-loss rates are slightly lower than mass-loss
rates determined mainly from UV line profiles \citep{bouhil,clres2}. On the
other hand, our predictions are a factor of about $ 4.7 $ lower than H$\alpha$
mass-loss rate estimates that were not corrected for clumping \citep[as compiled
by][]{mokz}. Our results can be reconciled with these H$\alpha$ estimates
assuming a clumping factor of at least
$C_\text{c}= 8 $.

Our predictions do not account for clumping. This is not a problem if the
clumping starts above the critical point, since the mass-loss rate is determined
below the critical point. However, if clumping starts relatively close to the
star \citep{clres2}, then the neglect of clumping may lead to the
underestimation of the predicted mass-loss rates. Consequently, the actual
mass-loss rates may be slightly higher \citep{sanya,muij}. The higher mass-loss
rates may be also supported by the fact that the analysis of \citet{clres2} that
accounts for optically thin and optically thick clumps gives systematically
higher mass-loss rates than predicted by us.

Our models still used the Sobolev approximation to calculate the radiative
bound-bound rates in the kinetic equilibrium equations. However, our test
calculations of bound-bound rates using CMF mean intensity showed that for many
of these rates the Sobolev approximation provides reliable estimate of the
radiation field mean intensity in the wind. Consequently, we expect that the
usage of Sobolev approximation to calculate the bound-bound rates has no
significant effect on the final model. On the other hand, the saving in
computing time is considerable.

From the discussion above it is not surprising that different codes predict
different ionization fluxes for $\lambda<400\,$\AA\ as found, for example, by
\citet{puluni}. The flux in this region is formed in the wind and is therefore
sensitive to the detailed density structure of the adopted model \citep[see
also][]{ssh}. This is not derived consistently in many models, which may be one
of the reasons for the difference of emergent fluxes derived from individual
models in the far UV region.

We have shown that global wind models predict wind mass-loss rate that agree
with empirical estimations (that were corrected for microclumping) within a
factor of about 1.6. This result shows a promising perspective for reconciling
all types of mass-loss rate determination in hot stars.

\begin{acknowledgements}
This work was supported by grant GA \v{C}R 13-10589S.
\end{acknowledgements}

\onecolumn
\appendix

\end{document}